\documentclass[12pt]{JHEP3}
\usepackage{amsfonts,amssymb}
\usepackage{amsmath}
\usepackage{color}
\usepackage{epsf,graphicx}

\title{ Aspects of ABJM orbifolds with discrete torsion}
\author{Mauricio Romo \\
 Department of Physics, UCSB, Santa Barbara, CA 93106}

\abstract{We analyze orbifolds with discrete torsion of the ABJM theory by a finite subgroup $\Gamma$ of $SU(2)\times SU(2)$ . Discrete torsion is
implemented by twisting the crossed product algebra resulting after orbifolding. It is shown that, in general, the order $m$ of the cocycle we chose to twist the algebra by enters in a non trivial way in the moduli space. To be precise, the M-theory fiber is multiplied by a factor of $m$ in addition to the other effects that were found before in the literature. Therefore we got a $\mathbb{Z}_{\frac{k|\Gamma|}{m}}$ action on the fiber. We present a general analysis on how this quotient arises along with a detailed analysis of the cases where $\Gamma$ is abelian.}
\begin{document}





\section{Introduction}

In the last two years some great amount of progress has been made towards a better understanding of the dynamics of M-theory and the $AdS_{4}/CFT_{3}$ correspondence. A  major step towards the formulation of a M-brane worldvolume action was the model proposed in the works by Bagger, Lambert and Gustavsson \cite{Bagger:2007jr,Bagger:2007vi,Gustavsson:2007vu}, the so called BLG theory, describing a stack of two coincident M2-branes probing a `M-fold' singularity \cite{Distler:2008mk}, whose interpretation is not clear yet. A complete non-abelian formulation came later, the so called ABJM theory \cite{Aharony:2008ug,Aharony:2008gk}. This theory describes a stack of $N$ M2-branes transversal to a $\mathbb{C}^{4}/\mathbb{Z}_{k}$ singularity, which corresponds to M-theory on $AdS_{4}\times S^{7}/\mathbb{Z}_{k}$. The theory is a Chern-Simons (CS) gauge theory with gauge group $G=U(N)_{k}\times U(N)_{-k}$, with $k, -k$ being the corresponding CS levels. The $\mathbb{Z}_{k}$ orbifold projects out four of the supercharges when $k>2$ so the supersymmetry is broken to $\mathcal{N}=6$ in three dimensions. The theory admits a perturbative expansion in a 't Hooft-like parameter $\lambda=\frac{N}{k}$, hence, for $k=1$ or $k=2$ (where the supersymmetry should get enhanced to $\mathcal{N}=8$), is strongly coupled.\\

One of the logical steps one can take to test this correspondence is to modify the $S^{7}/\mathbb{Z}_{k}$ internal space, for example, by a marginal deformation of ABJM or an orbifold of the gauge theory. Some of the simplest modifications of the background we can study are certain families of orbifolds. In principle we can consider orbifolds by any discrete subgroup of $SU(4)$. However $\mathcal{N}=1$ supersymmetric theories in three dimensions are non holomorphic, so in order to keep a better control of the resulting theory, we will consider orbifolds of the form $S^{7}/\Gamma$ where $\Gamma\subseteq SU(2)\times SU(2)$ breaking the supersymmetry at most to $\mathcal{N}=2$ in the three dimensions. This also allows us to work in a convenient $\mathcal{N}=1$ four dimensional superfield formulation. Orbifolds of this type, for $\Gamma$ any A-D-E group, have been already considered \cite{Hanany:2008qc,Berenstein:2009ay} and also many toric setups (see for example \cite{Hanany:2008fj,Imamura:2008qs,Franco:2008um,Hanany:2008gx,Davey:2009sr} and references therein) along with some non-toric deformations \cite{Martelli:2009ga,Forcella:2009jj}. One case that have been missed is the inclusion of discrete torsion in the orbifolds. The purpose of this paper is to study the role of discrete torsion in the cases $\Gamma\subseteq SU(2)\times SU(2)$. Our objective is to analyze carefully the moduli space resulting from placing M2-branes in these types of singularities. We will see that, as it happens in the case of orbifold singularities, the behavior of M2-branes does not mimic their D-brane counterparts \cite{Berenstein:2008dc,Benna:2008zy,Terashima:2008ba}.\\

Discrete torsion in string theory was originally considered by Vafa \cite{Vafa:1986wx}, for closed strings on orbifolds $X/\Gamma$. In that paper was shown that the partition function  $\sum Z_{g,g'}$ at one loop admits the insertion of phases $\varepsilon(g_{i},g_{j})$ multiplying the twisted sectors $Z_{(g_{i},g_{j})}$ and still preserving modular invariance. Modular invariance for genus $g>1$ surfaces impose extra conditions on $\varepsilon(g_{i},g_{j})$ that allows to write them as
$$
\varepsilon(g_{i},g_{j})=\frac{\alpha(g_{i},g_{j})}{\alpha(g_{j},g_{i})}\qquad [\alpha]\in H^{2}(\Gamma,U(1)).
$$
From the point of view of the states, the inclusion of these phases is equivalent to imposing the physical state condition $g\cdot| s\rangle=\varepsilon(h,g)| s\rangle$ on the states $| s\rangle$ in the sector twisted by $h$. On the other hand, the way this is reflected in the open string spectrum is that, invariance of the OPEs under $\Gamma$ \cite{Douglas:1998xa,Douglas:1999hq,Gomis:2000ej,Diaconescu:1997br} enforces that $\Gamma$ acts on the Chan-Paton factors by a projective representation with cocycle $\tilde{\alpha}\in [\alpha]$. If we want to compute the moduli spaces of these orbifold theories, this is obviously an effect we have to take into account. Our approach will be to construct an orbifold gauge theory from ABJM using the orbifold construction \cite{Douglas:1996sw,Lawrence:1998ja} that is well known for D-branes. After doing that we will obtain the moduli space by computing the chiral ring. In order to include discrete torsion we are then instructed to perform the orbifold projection using projective representations. As proposed in \cite{Berenstein:2001jr}, branes are inherently non-commutative objects and therefore can be represented by matrices that endow a representation of some algebra $\mathcal{A}$ derived from the F-term and D-term relations. When the transverse space to the branes is an orbifold $X/\Gamma$ the algebra $\mathcal{A}$ can naturally be associated with a crossed product algebra $\mathcal{A'}\boxtimes \Gamma$ in a way we will explain later. Then, the inclusion of discrete torsion on $X/\Gamma$ ultimately should be connected with the twisting of $\mathcal{A'}\boxtimes \Gamma$ by a cocycle $\alpha$. This is not direct since the path algebra $\mathbb{C}Q_{\Gamma}$ \cite{Berenstein:2002ge} derived from the orbifold construction \cite{Douglas:1996sw,Lawrence:1998ja} is not isomorphic to $\mathcal{A'}\boxtimes \Gamma$ in general. When the cocycle is trivial and $\Gamma$ is abelian the isomorphism holds, but in any other cases it does not. However, they have been proven to be Morita equivalent in a wide variety of cases \cite{Ginzburg:2006fu}. This is the property that will help us in our analysis since it implies that both algebras share characteristics that are relevant to compute physical quantities, such as the parameter space of simple modules or the center. Our interest in these objects relies in the fact that they are key in computing the chiral ring or the moduli space of vacua \cite{Berenstein:2001jr}.\\

This algebraic approach is the one we will take in this paper, since it seems to be the more convenient for computations in our cases and for possible generalizations of the results. However, twisting the algebra and finding the simple modules is not the whole story. One of the new ingredients that M2-branes will provide is the appearance of non-perturbative operators in the chiral ring. Upon compactification on a $S^{1}$, M-theory is dual to Type IIA and therefore the worldvolume theory of M2-branes should correspond to its counterpart for D2-branes (three dimensional SYM) after flow to the IR fixed point. The IIA picture of this duality can be seen by writing $S^{7}$ as a Hopf fibration of $\mathbb{CP}^{3}$
$$
 \begin {array}{ccc}S^{1}&\longrightarrow & S^{7}
 \\\noalign{\medskip}\text{\ \ }&\text{\ \ }&\downarrow
\\\noalign{\medskip}\text{\ \ }&\text{\ \ }&\mathbb{CP}^{3}\end {array}
$$

The $\mathbb{Z}_{k}$ orbifold acts just on the $S^{1}$ fiber. Therefore we can think of ABJM theory as dual to type IIA on $AdS_{4}\times \mathbb{CP}^{3}$ with RR 2-form flux turned on, which corresponds to the curvature of the connection on the circle bundle. This picture has been shown to persist in more general setups \cite{Aganagic:2009zk}. For instance, a toric Calabi-Yau (CY) fourfold $X_{4}$ can be written as a circle fibration over a $\mathbb{R}\times X_{3}$ base, where $X_{3}$ is a CY threefold. Therefore, M-theory in $X_{4}$ is dual to IIA in $X_{3}\times \mathbb{R}$ plus fluxes. The RR fluxes induce CS terms in the worldvolume theory of the D2-branes and the levels correspond to D6 and D4 brane charges of branes wrapping vanishing cycles on the $X_{3}$ \cite{Aganagic:2009zk}. Then we expect that in general, changing the CS levels will change the three dimensional moduli space. For example, in the simplest cases of orbifolds, performing the projection, as we mention before, will rescale the CS levels. This cannot be undone by rescaling the fields and is reflected in the moduli space one obtains. The way these effects show up is because of the existence of non-perturbative BPS operators, the so called monopole operators. They are therefore essential for the analysis of these families of theories (see for instance \cite{Kim:2010ac}, for a very recent work in the subject, and references therein). For instance, the enhancement of supersymmetry expected when $k=1,2$ has actually been proved to occur and monopole operators provide the necessary mechanism for it \cite{Samtleben:2010eu,Benna:2009xd,Bashkirov:2010kz,Gustavsson:2009pm,Kwon:2009ar}. Extra gauge invariant operators of dimension $1$ can be built by pairing monopole operators with scalars only when $k=1,2$ (since the dimensions of monopole operators depend on $k$) providing us with the missing conserved currents. In conclusion, monopole operators play an important role in the theories derived from M2-branes and their incorporation into the algebraic framework previously mentioned is one of the challenges we face.\\

The paper is organized as follows. Sections 2,3 and 4 are mainly review sections. In section 2 we review the ABJM theory and set up some notation and also review the semiclassical techniques we will use to compute the spectrum of monopole operators. In section 3 we review the orbifold projection for D-branes and projective representations in this context as well. Section 4 is devoted to presenting the mathematical background that will be useful in computing the simple modules and other objects we will need in our analysis. In section 5 we apply these mathematical tools for the particular case of orbifolds of ABJM showing that the M-theory fiber is multiplied by a factor of the order of the cocycle that characterizes the discrete torsion. Sections 6 and 7 are a detailed analysis of the case of abelian orbifolds. These cases are quite interesting because the Schur multiplier, $H^{2}(\Gamma, U(1))$ is larger. In section 8 we comment on the possible gravity duals to the theories previously analyzed. Finally, in section 9 we present some conclusions and possible future directions of this work. Complementary results are collected in the appendices.

\section{SCFTs in 3d and ABJM}\label{chiralsec}

\subsection{Review of ABJM}

We begin by briefly reviewing ABJM theory following  \cite{Benna:2008zy}, in order to set the conventions for the rest of the paper. This theory is better formulated in the language of $\mathcal{N}=1$ superfields inherited from 4d. So, let $\theta_{\alpha}$, $\bar{\theta}_{\alpha}$ be the complex Grassman numbers parametrizing the superspace and proceed as usual. Their indices are raised and lowered with $\varepsilon_{\alpha\beta}$. Let begin by looking at the vector multiplet. In the Wess-Zumino (WZ) gauge, it has the form
\begin{eqnarray}
V=2i\bar{\theta}\theta\sigma+2\theta\sigma^{\mu}\bar{\theta}A_{\mu}+i\sqrt{2}\theta\theta\bar{\theta}\chi^{\dag}-i\sqrt{2}\bar{\theta}\bar{\theta}\theta\chi+\theta\theta\bar{\theta}\bar{\theta}D.
\end{eqnarray}
Note that here, the auxiliary real scalar field $\sigma$ cannot be gauged away (it can be seen as the dimensional reduction of $A_{3}$). The group indices are omitted ($V=V^{a}T^{a}$ with $T^{a}$ the generators of $Lie(G)$). The action for the CS term is given by
\begin{eqnarray}
S_{CS}=-iK\int d^{3}xd^{4}\theta\int_{0}^{1}dtTr\Big[V\overline{D}^{\alpha}\Big(e^{tV}D_{\alpha}e^{-tV}\Big)\Big],
\end{eqnarray}
with $K=\frac{\kappa}{8\pi}$, where $\kappa$ is the so called CS level and is quantized. The derivatives $D$ and $\overline{D}$ are the covariant superspace derivatives. The other piece we will need is the matter field content. These are bi-fundamental fields. The gauge group is given by $G=U(N)\times U(\overline{N})$ (In \cite{Aharony:2008ug} $N=\overline{N}$, but if we include fractional M2-branes we can take any $N,\overline{N}$ \cite{Aharony:2008gk}). We will denote $\mathcal{Z}^{a}_{\hat{a}}$ and $\mathcal{W}^{\hat{a}}_{a}$ a superfield transforming in the representations $(\Box,\overline{\Box})$ and $(\overline{\Box},\Box)$ of $G$ respectively
\begin{eqnarray}
\mathcal{Z}\rightarrow U\mathcal{Z}\hat{U}^{\dagger},\nonumber\\
\mathcal{W}\rightarrow \hat{U}\mathcal{W}U^{\dagger}.
\end{eqnarray}
For a theory with two chiral fields $\mathcal{Z}$ and $\mathcal{W}$ and two anti-chiral fields $\bar{\mathcal{Z}}$ and $\bar{\mathcal{W}}$ transforming under $G$ as
\begin{center}
\begin{tabular}{||c|c|c||}
\hline
Field & $U(N)$ & $U(\overline{N})$ \\ \hline\hline%
$\mathcal{Z}$, $\bar{\mathcal{W}}$ & $\Box$ & $\overline{\Box}$ \\ \hline%
$\mathcal{W}$, $\bar{\mathcal{Z}}$ & $\overline{\Box}$ & $\Box$ \\
\hline\hline
\end{tabular}
\end{center}
the canonical kinetic term is given by
\begin{eqnarray}
S_{kin}=\int d^{3}xd^{4}\theta Tr\Big[-\bar{\mathcal{Z}}e^{-V}\mathcal{Z}e^{\hat{V}}-\bar{\mathcal{W}}e^{-\hat{V}}\mathcal{W}e^{V}\Big],
\end{eqnarray}
where $\hat{V}$ is the vector multiplet corresponding to the connection for $U(\overline{N})$.\\
The ABJM theory has two pair of these fields, say $\mathcal{Z}^{A}$, $\bar{\mathcal{W}}^{A}$, $\mathcal{W}_{A}$, $\bar{\mathcal{Z}}_{A}$ with $A=1,2$ and a superpotential term
\begin{eqnarray}
S_{pot}=\frac{1}{4K}\int d^{3}xd^{2}\theta Tr\Big[\varepsilon_{AC}\varepsilon^{BD}\mathcal{Z}^{A}\mathcal{W}_{B}\mathcal{Z}^{C}\mathcal{W}_{D}\Big]+\frac{1}{4K}\int d^{3}xd^{2}\theta Tr\Big[\varepsilon^{AC}\varepsilon_{BD}\bar{\mathcal{Z}}_{A}\bar{\mathcal{W}}^{B}\bar{\mathcal{Z}}_{C}\bar{\mathcal{W}}^{D}\Big].
\end{eqnarray}
which correspond to the conifold superpotential. To get a better insight of the theory is helpful to look at expressions in terms of components fields. The chiral fields components are given by
\begin{eqnarray}
\mathcal{Z}^{A}(x_{L})=Z^{A}(x_{L})+\sqrt{2}\theta \zeta^{A}(x_{L})+\theta^{2}F^{A}(x_{L}),\nonumber\\
\mathcal{W}_{A}(x_{L})=W_{A}(x_{L})+\sqrt{2}\theta \omega_{A}(x_{L})+\theta^{2}G_{A}(x_{L}),\nonumber\\
\bar{\mathcal{Z}}_{A}(x_{R})=Z^{\dagger}_{A}(x_{R})-\sqrt{2}\bar{\theta} \zeta^{\dagger}_{A}(x_{R})-\bar{\theta}^{2}F_{A}(x_{R}),\nonumber\\
\bar{\mathcal{W}}^{A}(x_{R})=W^{\dagger A}(x_{R})-\sqrt{2}\bar{\theta} \omega^{\dagger A}(x_{R})-\bar{\theta}^{2}G^{\dagger A}(x_{R}).
\end{eqnarray}
with $x_{L}^{\mu}=x^{\mu}+i\theta\sigma^{\mu}\bar{\theta}$, $x_{R}^{\mu}=x^{\mu}-i\theta\sigma^{\mu}\bar{\theta}$. As fermions will not play any role in our analysis we will omit them in the following to keep the formulas more clear. The action in term of the component fields is
\begin{eqnarray}
S_{ABJM} &=&\int d^{3}x\left[ 2K\varepsilon^{\mu\nu\lambda}Tr\Big(A_{\mu}\partial_{\nu}A_{\lambda}+\frac{2i}{3}A_{\mu}A_{\nu}A_{\lambda}-\hat{A}_{\mu}\partial_{\nu}\hat{A}_{\lambda}-\frac{2i}{3}\hat{A}_{\mu}\hat{A}_{\nu}\hat{A}_{\lambda}\Big)-4KD\sigma+4K\hat{D}\hat{\sigma}\right]\nonumber\\
&+&\int d^{3}x\Big[ -Tr(\mathcal{D}_{\mu}Z)^{\dag}_{A}\mathcal{D}^{\mu}Z^{A}-Tr(\mathcal{D}_{\mu}W)^{\dag A}\mathcal{D}^{\mu}W_{A}+W^{\dag}(\hat{D}W-WD)+Z^{\dag}(DZ-Z\hat{D})\nonumber\\
&+&|\hat{\sigma}W-W\sigma|^{2}+|\sigma Z-Z\hat{\sigma}|^{2}+ G^{\dag}G+F^{\dag}F\Big]\nonumber\\
&+&\frac{1}{K}\int d^{3}x\Big[\varepsilon_{AC}\varepsilon^{BD}(2F^{A}W_{B}Z^{C}W_{D}+2Z^{A}W_{B}Z^{C}G_{D})\nonumber\\
&-&\varepsilon^{AC}\varepsilon_{BD}(2F^{\dag}_{A}W^{\dag B}Z^{\dag}_{C}W^{\dag D}+2Z^{\dag}_{A}W^{\dag B}Z^{\dag}_{C}G^{\dag D})\Big],
\end{eqnarray}
solving for the auxiliary fields gives
\begin{eqnarray}\label{auxfields}
F^{\dag}_{A}=-\frac{1}{2K}\varepsilon_{AC}\varepsilon^{BD}W_{B}Z^{C}W_{D}\nonumber\\
G^{\dag A}=\frac{1}{2K}\varepsilon^{AC}\varepsilon_{BD}Z^{B}W_{C}Z^{D}\nonumber\\
\hat{\sigma}^{a}(\hat{T}^{a})^{\hat{i}}_{\ \ \hat{j}}=\frac{1}{4K}(Z^{\dag}Z-WW^{\dag})^{\hat{i}}_{\ \ \hat{j}}\nonumber\\
\sigma^{a}(T^{a})^{i}_{\ \ j}=\frac{1}{4K}(ZZ^{\dag}-W^{\dag}W)^{i}_{\ \ j}
\end{eqnarray}
and the gauge covariant derivative is given by
\begin{eqnarray}
\mathcal{D}_{\mu}W=\partial_{\mu}W-iWA_{\mu}+i\hat{A}_{\mu}W,\nonumber\\
\mathcal{D}_{\mu}Z=\partial_{\mu}Z+iA_{\mu}Z-iZ\hat{A}_{\mu}.
\end{eqnarray}
To finish, we recall the vacuum equations
\begin{eqnarray}
F=G=0,
\end{eqnarray}
\begin{eqnarray}\label{dterm}
\sigma Z-Z\hat{\sigma}=0,\nonumber\\
\hat{\sigma}W-W\sigma=0.
\end{eqnarray}

\subsection{BPS states and the chiral ring}
In this section we will review the characterization of the moduli space of vacua of SUSY field theories via the chiral ring operators (see \cite{Cachazo:2002ry} for details) and focus in particular on
3d SCFTs that have CS terms. The moduli space of vacua of SUSY gauge theories can be described in terms of expectation values of scalar gauge invariant operators $\mathcal{O}(\theta,\bar{\theta},x)$ in the chiral ring. These operators satisfy
\begin{eqnarray}
\overline{D}_{\alpha}\mathcal{O}(\theta,\bar{\theta},x)=0,
\end{eqnarray}
and their expectation values form a ring, as well
\begin{eqnarray}
\partial_{x_{1}}\langle\mathcal{O}(x_{1})\mathcal{O}(x_{2})\rangle &=&\partial_{x_{2}}\langle\mathcal{O}(x_{1})\mathcal{O}(x_{2})\rangle=0,\nonumber\\
\langle\mathcal{O}(x_{1})\mathcal{O}(x_{2})\rangle &=&\langle\mathcal{O}(x_{1})\rangle\langle\mathcal{O}(x_{2})\rangle,
\end{eqnarray}
so the chiral ring operators can be defined as the set
\begin{eqnarray}
\mathfrak{R}=\left\{\mathcal{O}|\overline{D}_{\alpha}\mathcal{O}(\theta,\bar{\theta},x)=0\right\}\Big/\left\{\mathcal{O}=\left\{\overline{D},G(\theta,\bar{\theta},x)\right\}\right\},
\end{eqnarray}
For SCFTs on the cylinder $\mathbb{R}\times S^{d-1}$ (that can be achieved via a Weyl rescaling of the metric) additional constrains can be imposed over the operators on $\mathfrak{R}$ due to the large amount of (super-)symmetry \cite{Berenstein:2007wi}. This boils down to consider operators whose lowest component $\phi$ is a superprimary in the chiral ring (i.e. its equivalence class can be represented by a superprimary). More importantly this casts $\phi$ as a BPS state satisfying $\Delta_{\phi}\sim R_{\phi}$, with $\Delta_{\phi}$ the scaling dimension of $\phi$ and $R_{\phi}$ its R-charge. In particular, for $d=3$
\begin{eqnarray}\label{bpseq}
\Delta_{\phi}= R_{\phi}.
\end{eqnarray}
So, the moduli space of these theories can be written as
\begin{eqnarray}
\mathcal{M}\cong \left\{\langle\phi\rangle| \mathcal{O}=\phi+\bar{\theta}\psi+\ldots,\mathcal{O}\in\mathfrak{R}\right\},
\end{eqnarray}
A proposal made by Berenstein \cite{Berenstein:2007wi,Berenstein:2005aa} suggests that the operators $\phi$  are in 1-1 correspondence with classical solutions of the equations of motion and the classical BPS equations. This means that the chiral ring operators should provide a holomorphic quantization of the space of these classical solutions.\\

Solving the classical equations is easier if we do radial quantization in $S^{d-1}\times \mathbb{R}$. Then, the Hamiltonian is equal to $\Delta$, the generator of dilatation.\\

It is a well known fact that the ABJM theory and orbifolds of it posses non-perturbative operators, a fact that can be seen from the classical equations \cite{Berenstein:2009sa,Berenstein:2009ay}. These operators are BPS and contribute to the chiral ring, and hence to the description of the moduli space of vacua   $\mathcal{M}_{3d}$. Not taking into account these states will result on an incorrect $\mathcal{M}_{3d}$ (indeed a non-complex variety). Therefore the superpotential alone does not give us all the information (for more details on the appearance of these extra massless degrees of freedom see \cite{Martelli:2008si} or  \cite{Imamura:2009ur}).\\
The connection between the appearance of a massless monopole and the representation theory from the superpotential algebra is given by the BPS equations. First note that in an orbifold of ABJM the Hamiltonian for the scalars, in the cylinder $S^{2}\times \mathbb{R}$ always take the form
\begin{equation}\label{lagrangianscalar}
\int_{S^2}\Big( Tr(\Pi_{\phi} \Pi_{\phi^{\dag}})+Tr( \mathcal{D} \phi (\mathcal{D} \phi)^{\dag})+\frac{1}{4} Tr(\phi^{\dag} \phi)+ V_{D}+V_{F}   \Big)
\end{equation}
where $\mathcal{D} \phi$ is the gauge covariant derivative in the sphere. The terms $V_{D}$ are the analogous of the D-terms coming from the supersymmetric CS action and from the canonical K\"ahler potential \footnote{Since we will deal with marginal deformations of the theory and preserve $\mathcal{N}=2$ supersymmetry in three dimensions, this will guarantee that the K\"ahler potential only receives corrections that are irrelevant in the IR \cite{Gaiotto:2007qi}}. $V_{F}$ is the scalar contribution to the superpotential. The sum over the arrows of the quiver is implicit and the important point is that the first three terms in (\ref{lagrangianscalar}) does not mix different arrows, only the arrow with its conjugate. Then, the BPS equations are
\begin{equation}\label{HQR}
H=Q_{R}
\end{equation}
where $Q_{R}$ is the R-charge, that is given by
\begin{equation}
Q_{R} = \sum_{\phi}\int_{S^2}  Tr(\frac{i}{2} \Pi_{\phi} \phi -\frac {i}{2}\Pi_{\phi^{\dag}}\phi^{\dag})
\end{equation}
since all the scalar fields have R-charge $\frac{1}{2}$. This is preserved by the orbifolds we will consider. The $R$-charge of the bifundamentals will not be modified because we are guarantee to have a canonical K\"ahler potential as we pointed out before. The equation (\ref{HQR}) is classical and it will give a sum of squares that must vanish, resulting in the following equations \cite{Berenstein:2009sa,Berenstein:2009ay}
\begin{eqnarray}
\mathcal{D} \phi=0
\\
 W_\phi= [\sigma,\phi]=0\\
\Pi_{\phi^{\dag}} =\dot \phi = \frac i 2 \phi
\end{eqnarray}
In addition we have to complement this with spherical symmetry (which is the classical condition of being a scalar) and the equations of motion. In particular the Gauss' law constraint will give us a relation for the magnetic fluxes on $S^{2}$ of the form $\Phi\equiv \int_{S^{2}}F\sim\phi\Pi_{\phi}$. If we quantize the moduli space after solving the representation theory for the quiver, we will see that $\phi\Pi_{\phi}$ is proportional to the number operator, so this imposes constraints on the wave functions which are equivalent to discrete identifications on the coordinate ring.\\

\subsection{Magnetic monopoles}

Here we will give a more detailed view of the monopole equations, focusing in a three dimensional SCFT in $S^{2}\times \mathbb{R}$, with the general assumptions we made in the last section. We want to solve the classical equations of motion and find the solutions corresponding to the states of the chiral ring. As we saw before, states of the chiral ring satisfy $H=Q_{R}$ where $H$ is the Hamiltonian identified with the scaling dimension and $Q_{R}$ just the (classical) R-charge. To be more precise let us write the conjugate momenta to our canonical variables. For this purpose, we denote our fields as $\phi^{(ab)}_{I}$ for arrows going from $V^{(a)}$ to $V^{(b)}$.
\begin{eqnarray}
\Pi_{A_{0}}&=&\Pi_{\hat{A}_{0}}=0\qquad\Pi_{A^{(a)}_{i}}=-2K_{a}\varepsilon^{0ij}A^{(a)}_{j},\nonumber\\
\Pi_{\phi^{(ab)\dag}_{I}}&=&\dot{\phi}^{(ab)}_{I}+iA^{(a)}_{0}\phi^{(ab)}_{I}-i\phi^{(ab)}_{I}A^{(b)}_{0}
\end{eqnarray}
The expression for $Q_{R}$ in terms of the fields can be read in \cite{Berenstein:2009sa}. Classical solutions corresponding to operators of the chiral ring must be spherically symmetric to have zero angular momentum, as we mentioned before, is the classical condition of being a scalar, then
\begin{eqnarray}\label{spherical}
\mathcal{D}_{i}F^{(a)}_{\mu\nu}=0\qquad i=\theta,\varphi.
\end{eqnarray}
Is convenient to choose the gauge so that $F_{ij}$ is diagonal and $A_{0}=0$. This gives the following conditions
\begin{eqnarray}\label{bpsclas}
\nabla_{i}\phi^{(ab)}_{I}&=&0\nonumber\\
F&=&G=\sigma^{(a)} \phi^{(ab)}_{I}-\phi^{(ab)}_{I}\sigma^{(b)}=0,\nonumber\\
F^{(a)}_{\theta\varphi}&=&j(\phi^{(ab)}_{I}).
\end{eqnarray}
The last condition is the constraint equation imposed by $A_{0}$ with $j(\phi^{(ab)}_{I})$ the source (the equation of motion of $A_{0}$). By $\mathcal{D}_{i}F_{\mu\nu}=0$, we have $F_{0i}=0$ and by our choice of gauge, $A_{0}=0$. So, $F^{(a)}_{\varphi\theta}=\widetilde{\Phi}^{(a)}$, where $\widetilde{\Phi}^{(a)}$ is a diagonal constant matrix, by the conditions of fiber bundles on $S^{2}$ \cite{Atiyah:1982fa}. Define the magnetic fluxes $\Phi^{(a)}=\int_{S^{2}} \frac{\widetilde{\Phi}^{(a)}}{\sin\theta}$ on the sphere. If we have bifundamental matter charged under $V^(a)$ and $V^{(b)}$ then Dirac quantization conditions requires that the fluxes satisfy
\begin{eqnarray}
\Phi^{(a)}_{ii}-\Phi^{(b)}_{jj}\in \mathbb{Z}\qquad\forall i,j
\end{eqnarray}
and since we are considering bifundamental fields with zero angular momentum along $S^{2}$
\begin{eqnarray}
\Phi^{(a)}-\Phi^{(b)}=0
\end{eqnarray}
both conditions must be imposed if there is a non-oriented path in the quiver joining $V^{(a)}$ and $V^{(b)}$. In all the examples considered here we will have connected quivers with no disjoint pieces (at least away from the singularities), so, assuming that, we can denote $\Phi^{(a)}=\Phi$ for all $a$. We still have an important subtlety to take into account, pointed out in \cite{Berenstein:2009sa}. This is, the quantization can admit fractional fluxes if all the ranks of the vertices are equal, so summarizing
\begin{eqnarray}\label{quantcond}
\Phi_{ii}&\in& \mathbb{Z}\qquad\forall i\text{\ \ if\ \ }dim(V^{(a)})\neq dim(V^{(b)})\text{\ \ for some\ \ }a,b\nonumber\\
\Phi_{ii}&=& m_{i}+a\qquad m_{i}\in\mathbb{Z}\text{\ \ }a\in\mathbb{Q}\text{\ \ if\ \ }dim(V^{(a)})=dim(V^{(b)})\text{\ \ }\forall a,b
\end{eqnarray}
 In principle $a\in \mathbb{R}$ but the constraints from $A_{0}$ gives an equation of the form $\Phi\sim Q_{R}$ restricting its values to be rational. Finally we write explicitly the equation of motion for $A^{(a)}_{0}$ (in the gauge $A^{(a)}_{0}=0$)
\begin{eqnarray}
-\frac{\kappa^{(a)}}{\pi\sin \theta}F^{(a)}_{\theta\varphi}&=&-i\sum_{I,b}\phi^{(ab)}_{I}\dot{\phi^{(ab)}_{I}}^{\dag}+i\sum_{I,b}\dot{\phi^{(ba)}_{I}}^{\dag}\phi^{(ba)}_{I}+h.c.\nonumber\\
&=&-i\sum_{I,b}\phi^{(ab)}_{I}\Pi_{\phi^{(ab)\dag}_{I}}+i\sum_{I,b}\Pi_{\phi^{(ba)}_{I}}\phi^{(ba)}_{I}+h.c.
\end{eqnarray}
integrating it we get
\begin{eqnarray}
-\kappa^{(a)}\Phi^{(a)}&=&\int_{S^{2}}\left(-i\sum_{I,b}\phi^{(ab)}_{I}\Pi_{\phi^{(ab)\dag}_{I}}+i\sum_{I,b}\Pi_{\phi^{(ba)}_{I}}\phi^{(ba)}_{I}+h.c.\right)
\end{eqnarray}

\section{The Orbifold projection}\label{orbp}

In this section we review the well known orbifold construction for D-branes \cite{Douglas:1996sw,Lawrence:1998ja} and how to introduce discrete torsion \cite{Douglas:1998xa,Douglas:1999hq} on them. This will give us the guidelines to construct our orbifold $3d$ gauge theory. As we will see in the following chapters, the moduli space of the orbifolded gauge theory is not the same as in the D-brane case which is natural, since we are dealing with M2-branes after all, not D-branes.\\

Consider an orbifold of a space $X$, of the form $X/\Gamma$ with $\Gamma$ a discrete group. When we place Dp-branes transversal to  $X$, the dual geometry in the near horizon limit will take the form $AdS\times Y$ where $Y$ is a compact variety and the real cone over $Y$ is isomorphic to $X$. $\Gamma$ is taken to be a discrete subgroup of $G_{R}$, the group of global symmetries of $Y$. The worldvolume theory we obtain on the branes are the well known quiver gauge theories that can be constructed by the orbifold projection prescription, proposed by Douglas and Moore \cite{Douglas:1996sw}. These theories are SCFTs whose R-symmetry group, $G_{R}$, is broken by the orbifold action and their field content and superpotential can be derived from the compatibility conditions applied to the fields of the unorbifolded theory. Let review this construction. First, we have to specify how we embed $\Gamma\leq G_{R}$. We also must choose a way in which $\Gamma$ acts on the Chan-Paton factors i.e. an embedding of $\Gamma$ in the gauge group $G=\prod_{a}U(N_{a})$. Denote the irreducible representations of $\Gamma$ by $\{R_{i}\}$, $i=1,\ldots, r$.\\

The embedding on $G$ must be of the form $\bigoplus_{i}n_{i}R_{i}$, for each vertex of the quiver associated to $G$ (we assume $\Gamma$ acts on $G$ without interchanging the nodes), with $n_{i}$ the multiplicities of each irreducible representation, this is usually denoted by
\begin{eqnarray}
\gamma=\bigoplus_{i}\mathbb{C}^{n_{i}}R_{i}
\end{eqnarray}
and we have to satisfy the constraint $N_{a}=\Sigma_{i}n_{i}dim(R_{i})$. We chose the action of $\Gamma$ in the points $x\in X$ to be proper, that is, the orbit of a generic point $x$ has $|\Gamma|$ distinct points (fixed points form closed subsets) then for a generic brane $i$ at a position $x(i)$ the action of $\Gamma$ will generate $|\Gamma|$ image branes $x(\gamma(g)(i))$ and so $\gamma(g)\in G$ will correspond to permutation matrices, that is, the regular representation. Therefore $n_{i}=dim(R_{i})$ and $N_{a}=|\Gamma|$.\\
The consistency conditions on the fields are the following. $A_{\mu}$ should be invariant under the orbifold
\begin{eqnarray}
\gamma(g)^{-1}A_{\mu}\gamma(g)=A_{\mu}\qquad \forall g \in \Gamma
\end{eqnarray}
this means that $A_{\mu}$ should be in the commutant of $\Gamma$. Since $A\in Hom(\mathbb{C}^{N},\mathbb{C}^{N})=\mathbb{C}^{N}\otimes(\mathbb{C}^{N})^{*}$, its invariant part under $\Gamma$ is $\bigoplus_{i}\mathbb{C}^{n_{i}}\otimes(\mathbb{C}^{n_{i}})^{*}\otimes\mathbf{1}_{n_{i}\times n_{i}}$ and $G$ gets broken accordingly
\begin{eqnarray}
A_{\mu}\rightarrow \bigoplus_{i}A_{\mu}^{(i)}\otimes\mathbf{1}_{n_{i}\times n_{i}}\qquad G=U(N)\rightarrow\prod_{i} U(n_{i}).
\end{eqnarray}
So, we have a new quiver where the nodes of the original quiver splits in vertices labeled by $i=1,\ldots,r$, and we have a gauge group $U(n_{i})$ associated to each of them. The factor $\mathbf{1}_{n_{i}\times n_{i}}$ gives a $n_{i}$ coefficient in front of the action of $A^{(i)}_{\mu}$ after taking the trace.\\
Now let see how this construction works for the matter. Denote the embedding of $\Gamma$ on $G_{R}$ by $R$. We just saw that the nodes corresponds to irreducible representations of $\Gamma$. Then, the arrows between the nodes, which corresponds to the bifundamental fields, are given by the invariant part of $\phi_{ss'}^{a}$ under
\begin{eqnarray}
\phi_{ss'}^{a}\rightarrow R(g)_{\ b}^{a}\gamma(g)^{-1}\phi_{ss'}^{b}\gamma(g)\qquad\forall g \in \Gamma
\end{eqnarray}
likewise the case of the gauge fields (define the Clebsh-Gordan coefficients $a^{R}_{ij}$ by $R\otimes R_{i}=\oplus_{j}a_{ij}^{R}R{j}$) this can be computed and the arrows of the new quiver are in
\begin{eqnarray}
R\otimes Hom(\mathbb{C}^{N},\mathbb{C}^{N})^{\Gamma}=\bigoplus_{i,j}a^{R}_{ij}\mathbb{C}^{n_{i}}\otimes(\mathbb{C}^{n_{j}})^{*}\otimes\mathbf{1}_{n_{j}\times n_{j}}
\end{eqnarray}
The way we encode the discrete torsion is by noting that the representation $\gamma$ is defined up to a phase. This, plus associativity, implies that we can consider projective representations \cite{Karpilovsky}.\\

\begin{figure}[h]
\centering
\includegraphics[height=3cm]{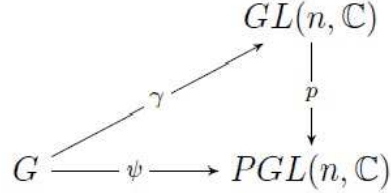}
\caption{Commutative diagram for projective representations}
\label{Fig.2}
\end{figure}

Projective representations are given by homomorphisms $\psi:\Gamma\rightarrow PGL(n,\mathbb{C})=GL(n,\mathbb{C})/\mathbb{C}^{*}$.The lift to $GL(n,\mathbb{C})$ is what we are interested in. Let $p:GL(n,\mathbb{C})\rightarrow PGL(n,\mathbb{C})$ be the canonical projection. Then we are looking for a lift $\gamma:\Gamma\rightarrow GL(n,\mathbb{C})$ such that $p\circ \gamma=\psi$ (Fig.\ref{Fig.2}).
Of course, $\gamma$ is not an homomorphism but it must satisfy the conditions
\begin{eqnarray}\label{eq1p}
\gamma(g)\gamma(h)=\alpha(g,h)\gamma(gh)\qquad \text{\ for all\ }g,h\in \Gamma,
\end{eqnarray}
where $\alpha:\Gamma\times G\rightarrow \mathbb{C}^{*}$ is called a factor set. Associativity imposes extra conditions on $\alpha$, say
\begin{eqnarray}\label{eq2p}
\alpha(g,h)\alpha(gh,k)=\alpha(h,k)\alpha(g,hk),
\end{eqnarray}
and we use the convention $\gamma(e)=1$, with $e$ the identity, so
\begin{eqnarray}
\alpha(g,e)=\alpha(e,g)=1\qquad \forall g\in \Gamma.
\end{eqnarray}
Two representations are say to be projectively equivalent if there exist a map $c:G\rightarrow \mathbb{C}^{*}$ such that
\begin{eqnarray}\label{eq3p}
\alpha(g,h)=c(g)c(h)c^{-1}(gh)\qquad \forall g,h\in \Gamma,
\end{eqnarray}
this is equivalent to say that we can make $\alpha=1$ by a rescaling of the elements of the group by a factor $c(g)$ plus a similarity transformation. The $\alpha$ maps are cocycles of $H^{2}(\Gamma,\mathbb{C}^{*})$ and if $\alpha$ satisfies (\ref{eq3p}) is called a coboundary. Moreover, as we are interested in projective representations of finite groups, by taking determinant on (\ref{eq1p}) we get
\begin{eqnarray}
\alpha(g,h)^{n}\det(gh)=\det(g)\det(h),
\end{eqnarray}
so taking $c(g)=det(g)$ allows us to redefine $\alpha$ up to a $n$th root of unity. This allows to replace $\mathbb{C}^{*}$ by $U(1)$ everywhere in our previous discussion. This also shows that for finite groups $H^{2}(\Gamma,U(1))$ must be finite and so the number of inequivalent projective representations.\\
Roughly speaking discrete torsion will appear then as extra phases in the superpotential \cite{Douglas:1998xa,Douglas:1999hq,Berenstein:2000hy}. This can be traced to the existence of a non-trivial background B-field along the transversal directions \cite{Seiberg:1999vs} and then, branes wrapping some 2-cycle $\Sigma$ on $X$ have their charge quantized depending on $\int_{\Sigma}B$.

\section{Mathematical setup}

In this section we will review the mathematical background that will be useful in our analysis. More details in some of the points discussed here can be found in \cite{Berenstein:2001jr,Berenstein:2002ge,homologicalDbranes}. In these references, this is treated in the context of D-branes, but what we will review here remains true in very general grounds. In the next section we will see what are the subtleties when dealing with the specific case of M2-branes.\\

Before we perform the orbifold projection, the fields $\phi_{I}$ will be represented by arrows of some quiver diagram. These arrows modulo the relations coming from the superpotential, $dW=0$, plus D-term relations, will span a $\mathbb{C}^{*}$-algebra \begin{eqnarray}
\mathcal{A}=\langle\phi_{I},P^{(a)}\rangle/\{V_{F}=0,V_{D}=0\},
\end{eqnarray}
where $P^{(a)}$ are the projectors associated with the vertices of the quiver. Upon orbifold projection we will obtain the path algebra $\mathbb{C}Q_{\Gamma}$. On the other hand, $\Gamma$ can be seen as an element of $Aut(\mathcal{A})$, that acts by conjugation
\begin{eqnarray}\label{orbconseq}
\phi_{I}^{g}=\gamma(g)\phi_{I}\gamma(g)^{-1}=R(g)_{I}^{\ J}\phi_{J}
\end{eqnarray}
$\Gamma$ and $\mathcal{A}$ together, along with the action of $\Gamma$ on $\mathcal{A}$, form a crossed product algebra $\mathcal{A}\boxtimes\Gamma$. An element $a\in \mathcal{A}\boxtimes\Gamma$ is written as
\begin{eqnarray}
a=\sum_{g\in\Gamma}a_{g}\rtimes e_{g}\qquad a_{g}\in \mathcal{A}, e_{g}\in \mathbb{C}\Gamma_{\alpha}
\end{eqnarray}
where $\mathbb{C}\Gamma_{\alpha}$ is the twisted group algebra of $\Gamma$ by the factor set (or cocycle) $[\alpha]\in H^{2}(\Gamma,U(1))$ defined by
\begin{eqnarray}
\mathbb{C}\Gamma_{\alpha}=\{ e_{g}| e_{g} e_{g'}=\alpha(g,g')e_{gg'}, g\in \Gamma\}
\end{eqnarray}
the multiplication rules of $\mathcal{A}\boxtimes\Gamma$ are given by
\begin{eqnarray}
(a\rtimes e_{g})(a'\rtimes e_{g'})=ae_{g}a'e_{g}^{-1}\rtimes e_{g}e_{g'}=\alpha(g,g')aa'^{g}\rtimes e_{gg'},
\end{eqnarray}
So, we can think of $\mathcal{A}\boxtimes\Gamma$ as a group algebra for $\Gamma$ with coefficients in $\mathcal{A}$. Suppose the simple modules of $\mathcal{A}$ are finite dimensional, to be more precise, suppose $\mathcal{A}$ is finitely generated as a module over its center. Then there exists a basis of generators $\{s_{a}\}$ such that we can write any element of $\mathcal{A}$ as
\begin{eqnarray}
\sum_{a}z_{a}s_{a}\qquad z_{a}\in \mathcal{ZA}
\end{eqnarray}
this implies that any simple module $\mu_{p}:\mathcal{A}\rightarrow M_{n}(\mathbb{C})$ can be characterized by a set of complex numbers $p$ which corresponds to the values of the generators of $\mathcal{ZA}$ (since they will be proportional to $\mathbf{1}_{n\times n}$) on that particular representation. Therefore, there exist a natural map $[\mu_{p}]\rightarrow \widetilde{X}$, where $[\mu_{p}]$ denote some equivalence class of modules, related by some subset of the similarity transformations $GL(n,\mathbb{C})$. The variety $\widetilde{X}$ is a commutative space, the space probed by the closed strings, and expected to be a covering of $X$.\\

The simple modules of $\mathcal{A}\boxtimes\Gamma$ can be constructed from those of $\mathcal{A}$. The action of $\Gamma$ on $\mathcal{A}$ induces a natural action of $g\in \Gamma$ on the modules $\mu_{p}$, say $e_{g}(\mu_{p})=\mu_{p^{g}}$. This gives the simple modules (they are simple by construction) of $\mathcal{A}\boxtimes\Gamma$ the form
\begin{eqnarray}
\bigoplus_{g\in\Gamma}\mu_{p^{g}}.
\end{eqnarray}
In this representation, the elements $e_{g}$ will be just permutation matrices, i.e., the regular representation, tensored with $\mathbf{1}_{n\times n}$. These modules are clearly unique when $p$ is a non singular point. When $p$ is held fixed by a subgroup $H\leq \Gamma$ then, more than one simple module can correspond to the same $p$ (they will be indeed classified by the irreducible representations of $H$). The dimension of these modules will be $|\Gamma|n$ if $p$ is a regular point i.e. $|Orb_{\Gamma}(p)|=|\Gamma|$, or, if $p$ gets fixed by some subgroup $H\lhd \Gamma$, then $|Orb_{\Gamma}(p)|=\frac{|\Gamma|}{|H|}$ and the dimension will be $\frac{|\Gamma|n}{|H|}$. Note that we did not mention the discrete torsion in the previous derivation. If we consider a non trivial cocycle, the representations will be exactly the same, with the regular representation for $\mathbb{C}\Gamma_{\alpha}$, which have the same form as in the case with no discrete torsion, say $\oplus_{i}dim(R_{i})R_{i}$ where the sum goes over all inequivalent projective representations with cocycle $\alpha$ \cite{Karpilovsky}.\\
Now, focus on the construction of the path algebra $\mathbb{C}Q_{\Gamma}$. The crossed product algebra is not expected to be isomorphic in a generic case, only Morita equivalent \footnote{For two algebras being Morita equivalent means that their categories of modules are equivalent. See for example \cite{Weibel}.}, which is enough for our purposes. In the case $\Gamma$ is abelian and $\alpha$ is trivial $\mathbb{C}Q_{\Gamma}\cong \mathcal{A}\boxtimes\Gamma$, but as long as $\alpha$ is non trivial this does not hold. For simplicity we will focus on the case of $\Gamma$ being abelian. Then, all projective representations for a given cocycle class $[\alpha]$ have the same dimension, say $s$. The number, $r$, of inequivalent projective representations is given by the number of $\alpha$-regular classes of $\Gamma$.\\

\section{M2-branes on orbifolds with discrete torsion}

In this section we will apply the mathematical formalism we reviewed in the last section to the particular case of M2-branes. We keep the same notation.
Denote the arrows of $\mathcal{A}$ by  $\phi^{(ab)}_{I}$, for bifundamentals joining the vertices $V^{(a)}$ and $V^{(b)}$, and the projectors by $P^{(a)}$, with $a,b=1,\ldots,Q_{0}$. The relations will be derived from a superpotential which is polynomial in the fields, schematically ($[l]$ denotes powers of the fields contracted in a gauge invariant way)
\begin{eqnarray}
W=Tr(\sum_{l}a_{[l]}\phi^{[l]}).
\end{eqnarray}
After performing the orbifold projection we described in section \ref{orbp} we will have arrows $\phi^{(ab)}_{I\ \ ij}$, joining the vertices $V^{(a)}_{i}\rightarrow V^{(b)}_{j}$ and projectors $P^{(a)}_{i}$ with $i,j=1,\ldots r$ which satisfy $P^{(a)}_{i}P^{(b)}_{j}=\delta^{ab}\delta_{ij}P^{(a)}_{i}$. In the abelian case we will have $\phi^{g}_{I}=\chi_{I}(g)\phi_{I}$ with $\chi_{I}(g)$ a character of $\Gamma$. The arrows of the projected quiver will be the Clebsch-Gordan coefficients of $\chi_{I}\otimes R_{i}=R_{\chi_{I}(i)}$. Let us say we fix a canonical form $R_{i}(g)\in M_{s}(\mathbb{C})$ for each projective irreducible representation. Then
\begin{eqnarray}
\chi(g)\cdot R_{i}(g)=U_{\chi(i)}R_{\chi(i)}(g)U^{\dag}_{\chi(i)}
\end{eqnarray}
where $U_{\chi(i)}\in U(s)$ is a matrix for the change of basis. Therefore, the superpotential for the projected fields will be given by $W$, but taking the trace over the fields $\phi^{(ab)}_{I\ \ i\chi_{I}(i)}\otimes U_{\chi_{I}(i)}$. The trace over the matrices $U_{\chi_{I}(i)}$ will insert phases between the terms in $W$. The kinetic terms will change just by a factor of $s$, they will be given by
\begin{eqnarray}
\mathcal{S}_{kin}=s\sum Tr\left(\mathcal{D}_{\mu}\phi^{(ab)}_{I\ \ ij}(\mathcal{D}^{\mu}\phi^{(ab)}_{I\ \ ij})^{\dag}\right)
\end{eqnarray}
and the CS levels will also be rescaled by $s$
\begin{eqnarray}
\kappa^{(a)}_{i}\rightarrow s\kappa^{(a)}_{i}
\end{eqnarray}
Note that we can rescale the bifundamental matter fields to have a canonical kinetic term, but we cannot get rid of the factor of $s$ in front of the CS levels. If we do this rescaling, the constraint equation will then read
\begin{eqnarray}\label{const1}
-\frac{s\kappa^{(a)}_{i}}{\pi\sin\theta}F^{(a)}_{\theta\varphi\ \ i}=-i\sum_{I,b,j}\phi^{(ab)}_{I\ \ ij}\dot{\phi^{(ab)}_{I\ \ ij}}^{\dag}+i\sum_{I,b,j}\dot{\phi^{(ba)}_{I\ \ ji}}^{\dag}\phi^{(ba)}_{I\ \ ji}+h.c.
\end{eqnarray}
Define the matrices $e_{ij}\in M_{r}(\mathbb{C})$ which have only a $1$ in the $ij$ position. Then, an obvious representation for the projected algebra is given by \footnote{We are not showing explicitly here that these modules are simple. However, arguments based on Morita equivalence shows this gives the right result. If $\mathbb{C}Q_{\Gamma}$ and $\mathcal{A}\boxtimes\Gamma$ are Morita equivalent (we will show it explicitly in our examples) , the parameters that describes their simple modules should be the same and the dimension of the simple $\mathbb{C}Q_{\Gamma}$-modules can be read form the explicit form of the correspondence.}
\begin{eqnarray}
R(\phi^{(ab)}_{I\ \ i\chi_{I}(i)})&=&\mu_{p}(\phi^{(ab)}_{I})\otimes e_{i\chi_{I}(i)}\otimes U_{\chi_{I}(i)}\nonumber\\
R(P^{(a)}_{i})&=&\mu_{p}(P^{(a)})\otimes e_{ii}\otimes \mathbf{1}_{s\times s}
\end{eqnarray}
and the gauge fields will take the form $A^{(a)}_{i}\rightarrow A^{(a)}\otimes e_{ii}\otimes \mathbf{1}_{s\times s}$. Therefore the constraint equation (\ref{const1}) keeps the same form but with the bifundamental fields replaced by $\mu_{p}(\phi^{(ab)}_{I})$, that is, basically the same constraint equation of the unorbifolded theory, but with the level rescaled by $s$. Then if we want to do holomorphic quantization on the variables $\phi^{(ab)}_{I}\equiv \mu_{p}(\phi^{(ab)}_{I})$ , we can replace this representation back in the action and the kinetic terms will be the same as of the original theory but rescaled by a factor of $rs$. Then the conjugated momenta associated with the $\phi^{(ab)}_{I}$ will be
\begin{eqnarray}
\Pi_{\phi^{(ab)}_{I}}=rs\dot{(\phi^{(ab)}_{I})}^{\dag}
\end{eqnarray}
and finally the constraint will be given by
\begin{eqnarray}\label{monopeq}
-\frac{s^{2}r\kappa^{(a)}}{2\pi}F^{(a)}_{\theta\varphi}=-\frac{|\Gamma|\kappa^{(a)}}{2\pi}F^{(a)}_{\theta\varphi}=-i\sum_{I,b}\phi^{(ab)}_{I}\Pi_{\phi^{(ab)\dag}_{I}}+i\sum_{I,b}\Pi_{\phi^{(ba)}_{I}}\phi^{(ba)}_{I}+h.c.
\end{eqnarray}
This factor of $|\Gamma|$ has appeared before (first noticed in \cite{Benna:2008zy,Terashima:2008ba} for the abelian case and then showed to hold for any A-D-E group in \cite{Berenstein:2009ay}) and its consequence is that the orbifold along the $S^{1}$ circle is not the naive one we expect , say $\mathbb{Z}_{\kappa}$, but instead $\mathbb{Z}_{|\Gamma|\kappa}$. When there is discrete torsion we have an additional effect. As usual, the CS levels are rescaled by $dim(R_{i})$, so $\kappa\rightarrow dim(R_{i})\kappa$. Let denote $e$, the exponent of $H^{2}(\Gamma, U(1))$, then a theorem by Schur says that $e^{2}\mid |\Gamma|$ and, moreover it can be shown that if $m[\alpha]=0$ then $m\mid dim(R_{i})$ for all $i$ (see appendix \ref{app:proof}) and so, the quantization condition on $\kappa$, in general will be $\kappa=\frac{k}{m}$,  $k\in \mathbb{Z}_{>0}$. In the particular case of ABJM, $\mathcal{A}$ is the conifold algebra $\mathcal{A}_{c}$ (see appendix \ref{app:conifold}) and the representations will be parametrized by four complex variables. The monomials that will represent the wave functions, after holomorphic quantization will be then of the form \cite{Berenstein:2009ay}
\begin{eqnarray}
f_{1}^{i_{1}}f_{2}^{i_{2}}g_{1}^{j_{1}}g_{2}^{j_{2}}
\end{eqnarray}
then, the gauge invariance condition, derived from the monopole equations (\ref{monopeq}) will be (note that the condition is more subtle than this, taking into account (\ref{quantcond}), but for now, we will consider the fractional piece of $\Phi$ to be zero)
\begin{eqnarray}
i_{1}-i_{2}+j_{1}-j_{2}\in \frac{|\Gamma|k}{m}\mathbb{Z}.
\end{eqnarray}
Then, the space probed by a D0-brane in the bulk \footnote{From the M-theory point of view, BPS monopoles are dual to D0-branes in type IIA string theory, with momentum in the extra fiber.} with  the will be given by $\mathbb{C}^{4}/(\mathbb{Z}_{\frac{|\Gamma|k}{m}}\times \Gamma)$. The singular locus, where the fractional branes are stuck, will admit a resolution by the algebra \cite{Berenstein:2001jr}. In this case, a point $p$ will be held fixed by a subgroup $H\leq \Gamma$. If $\Gamma$ is abelian, then $H$ is a normal subgroup and we can cast the orbifold as $X/\Gamma\rightarrow (X/H)/(\Gamma/H)$ \cite{Berenstein:2000mb}. The orbifold $X/H$ is trivial at these fixed points and so, it gives $r_{H}$ copies of $X$, with $r_{H}$ the number of irreducible representations ($H_{i}$) of $H$ with induced cocycle $\tilde{\alpha}$. There is again a rescaling of the CS levels $\kappa\rightarrow dim(H_{i})\kappa$. Then we will have $r_{H}$ copies of the orbifold $X/(\Gamma/H)$, each with levels $\kappa_{i}=dim(H_{i})\kappa$. Applying then the same prescription as before, the radius of the M-theory circle will be reduced by a factor of $\frac{dim(H_{i})|\Gamma|k}{|H|m}$ at each subquiver (note that the quantization condition on the CS levels must be inherited from the result we got in the bulk). Therefore the tension of a fractional brane will be given by $\frac{dim(H_{i})|\Gamma|k}{2|H|m}$. Adding them with the correct multiplicity gives
\begin{eqnarray}
T_{D0}=\sum_{i=1}^{r_{H}}\frac{dim(H_{i})^{2}|\Gamma|k}{2|H|m}=\frac{|\Gamma|k}{2m}
\end{eqnarray}
which is the expected result, for the tension of a D0-brane in the bulk.

\section{Example 1: $q$-deformed algebra $\mathcal{A}_{q}$}\label{sec:qdefsec}

In this section we will compute the simplest orbifold with non trivial discrete torsion in a detailed way. This corresponds to a $\mathbb{Z}_{n}\times \mathbb{Z}_{n}$ orbifold of the conifold with maximal discrete torsion i.e. the order of $[\alpha]$ is $n$. In the following section, the general case of a $\mathbb{Z}_{n}\times \mathbb{Z}_{m}$ orbifold will be analyzed, nevertheless is illustrative to go through the simplest one in detail first. We will begin computing the moduli space in the non-singular locus i.e. for a brane in the bulk or Azumaya locus of the superpotential algebra and we will focus on the singularities after that.

\subsection{Moduli space of a regular M2-brane in the bulk}\label{sec:regm2}

Consider a $q$-deformation of the ABJM superpotential
\begin{eqnarray}\label{conifoldq2}
W=Tr\left(A_{1}B_{1}A_{2}B_{2}-qA_{1}B_{2}A_{2}B_{1}\right),
\end{eqnarray}
where $q$ is a $n$th root of unity $q=e^{\frac{2\pi im}{n}}$ ($m$ and $n$ are relatively prime). This deformation will not modify the quiver, however, we will see that the moduli space we get from it, is the same as the one obtained from a $\mathbb{Z}_{n}\times\mathbb{Z}_{n}$ quotient with cocycle $[q]$. The algebra $\mathcal{A}_{q}$ of the $q$-deformed conifold, derived from (\ref{conifoldq2}) is spanned by $A_{i}$, $B_{i}$ and the projectors of the two nodes, $P_{1}$ and $P_{2}$. The path algebra relations are given by
\begin{eqnarray}
P_{1}A_{i}=A_{i}P_{2}\qquad P_{2}A_{i}=A_{i}P_{1}=0, \\\nonumber
P_{1}B_{i}=B_{i}P_{2}=0\qquad P_{2}B_{i}=B_{i}P_{1},
\end{eqnarray}
and the relations derived from (\ref{conifoldq2})
\begin{eqnarray}
A_{2}B_{2}A_{1}=qA_{1}B_{2}A_{2},\\\nonumber
A_{1}B_{1}A_{2}=qA_{2}B_{1}A_{1},\\\nonumber
B_{1}A_{2}B_{2}=qB_{2}A_{2}B_{1},\\\nonumber
B_{2}A_{1}B_{1}=qB_{1}A_{1}B_{2}.
\end{eqnarray}
Define $X=A_{1}+B_{1}$, $Y=A_{2}+B_{2}$, $\pi=P_{2}-P_{1}$ and $\Sigma=\sigma+\hat{\sigma}$, so the D-term equations (\ref{dterm}) can be written in a compact form
\begin{eqnarray}
[X,\Sigma]=[Y,\Sigma]=0\qquad \Sigma=\pi(XX^{\dag}+YY^{\dag}-X^{\dag}X-Y^{\dag}Y),
\end{eqnarray}
from the point of view of the algebra, the D-term conditions are essentially telling us to enlarge the center of $\mathcal{A}_{q}$ by $\Sigma\in \mathcal{ZA}_{q}$ and therefore if we are interested in simple modules of $\mathcal{A}_{q}$, $\Sigma$ must be proportional to the identity. In four dimensions the equation of $\Sigma$ will correspond to the familiar symplectic quotient \cite{Klebanov:1998hh}. For the unresolved conifold we will have $\Sigma=0$ and, as shown in \cite{Luty:1995sd}, we can ignore it and quotient by the complexified gauge group $G^{\mathbb{C}}$. If we include FI terms the singularity gets resolved and $\Sigma$ has a fixed value. In the present case, the equation of $\Sigma$ can be ignored too, but we should have in mind that there is a $U(1)$ subgroup corresponding to the dual photon becoming massless at the origin that we should not consider as a gauge symmetry and therefore we should not quotient by it a priori. An alternative way to see this, from a path algebra point of view is the following. If we have an $\mathcal{A}_{q}$-module, a similarity transformation will act as $X\rightarrow GXG^{-1}$ with $G$ of the form $G_{+}\oplus G_{-}$ and $G_{\pm}\in GL(N_{\pm},\mathbb{C})$ with $N_{\pm}$ the rank of the vertices of the module. However this transformation will act on $\Sigma$ as $\Sigma\rightarrow G\Sigma G^{\dag}$, then if we want it to be a similarity transformation we need to consider $G_{\pm}\in U(N_{\pm})$ instead.\\

Now, consider the simple modules of $\mathcal{A}_{q}$
\begin{equation}\label{solalg}
\pi=%
\begin{bmatrix}
\mathbf{1}_{n\times n} & 0  \\
0 & -\mathbf{1}_{n\times n}%
\end{bmatrix}%
\qquad Y=%
\begin{bmatrix}
0 & f_{2}P  \\
g_{2}I  & 0%
\end{bmatrix}%
\qquad X=%
\begin{bmatrix}
0 & f_{1}Q  \\
g_{1}Q^{-1}P^{-1}  & 0%
\end{bmatrix}%
\end{equation}
or, in terms of the $A$ and $B$ fields
\begin{eqnarray}
A_{2}=g_{2}I\qquad A_{1}=g_{1}Q^{-1}P^{-1}\qquad B_{2}=f_{2}P\qquad B_{1}=f_{1}Q
\end{eqnarray}
the matrices $P$ and $Q$ are given by
\begin{equation}\label{pqmatrices}
P=%
\begin{bmatrix}
1 & 0 & 0 &\cdots & 0 \\
0 & q & 0 & \cdots & 0 \\
0 & 0 & q^{2} & \cdots & 0 \\
0 & 0 & 0 & \ddots & 0 \\
0 & 0 & 0 & \cdots & q^{n-1} %
\end{bmatrix}%
\text{, \ }Q=%
\begin{bmatrix}
0 & 1 & 0 &\cdots & 0 \\
0 & 0 & 1 & \cdots & 0 \\
0 & 0 & 0 & \ddots & 0 \\
0 & 0 & 0 & \cdots & 1 \\
1 & 0 & 0 & \cdots & 0 %
\end{bmatrix}%
\text{,}
\end{equation}
a direct computation gives
\begin{equation}
\Sigma=%
\begin{bmatrix}
|f_{1}|^{2}+|f_{2}|^{2}-|g_{1}|^{2}-|g_{2}|^{2} & 0  \\
0 & |f_{1}|^{2}+|f_{2}|^{2}-|g_{1}|^{2}-|g_{2}|^{2}%
\end{bmatrix}%
\end{equation}
so, $\Sigma$ is proportional to the identity, as expected, and then we do not need to impose further conditions to satisfy (\ref{dterm}). Now, let examine how the remaining gauge transformations act. Since we do not want to introduce additional moduli, the transformations must have the form
\begin{equation}
U=%
\begin{bmatrix}
U_{1} & 0  \\
0  & \lambda U_{1}%
\end{bmatrix}%
\end{equation}
to preserve $A_{2}\sim 1$. In order to preserve $B_{2}\sim P$ is easy to show that $U_{1}$ must be of the form
\begin{eqnarray}
U_{1}=Q^{a}P^{b},
\end{eqnarray}
therefore, under these family of transformations, the matrix form of $X,Y$ does not change, only the factors $f_{i}$ and $g_{i}$ and they change in the following way
\begin{eqnarray}\label{gauget}
(f_{1},f_{2},g_{1},g_{2})\mapsto (\lambda^{-1}q^{-b}f_{1},\lambda^{-1} q^{a}f_{2},\lambda q^{b-a}g_{1},\lambda g_{2}).
\end{eqnarray}
This can be separated as the action of $\lambda$ and two independent $\mathbb{Z}_{n}$ factors on $\mathbb{C}^{4}$
\begin{eqnarray}\label{actionorb}
\lambda:(f_{1},f_{2},g_{1},g_{2})&\mapsto& (\lambda^{-1}f_{1},\lambda^{-1} f_{2},\lambda g_{1},\lambda g_{2})\nonumber\\
\tau_{1}:(f_{1},f_{2},g_{1},g_{2})&\mapsto& (q^{-1}f_{1},f_{2},qg_{1}, g_{2})\nonumber\\
\tau_{2}:(f_{1},f_{2},g_{1},g_{2})&\mapsto& (f_{1},qf_{2},q^{-1}g_{1},g_{2}).
\end{eqnarray}
Here is a good point to compare again with the $4d$ case. Being in $4d$ we would consider complexified gauge transformations $U\in GL(2n,\mathbb{C})$ and so $\lambda\in \mathbb{C}^{*}$ then the moduli space spanned by $\{f_{i},g_{i}\}$ will be $(\mathbb{C}^{4}/\mathbb{C}^{*})/\mathbb{Z}_{n}\times\mathbb{Z}_{n}$ which is exactly the orbifold of the conifold by $\mathbb{Z}_{n}\times\mathbb{Z}_{n}$. For the $3d$ case $U\in U(2n)$ hence $\lambda\in U(1)$. So if we naively quotient by the action of $\lambda$, locally, we will obtain $(\mathbb{C}^{3}\times \mathbb{R})/\mathbb{Z}_{n}\times\mathbb{Z}_{n}$ which is not even a complex manifold and so we know is the wrong answer for $\mathcal{M}_{3d}$. From the field theory point of view what is happening is that we are missing operators in the chiral ring: the monopoles, as mentioned before.\\
So, we regard the space we found, prior to quotient by $U(1)$, $\mathbb{C}^{4}/\mathbb{Z}_{n}\times\mathbb{Z}_{n}$, as a covering of the moduli space, and there must be a residual action of a $\mathbb{Z}_{l}\subset U(1)$ after we take into account the monopoles.\\
We now use the prescription we reviewed in section \ref{chiralsec}. First we replace our solution on the Lagrangian (\ref{lagrangianscalar}) to holomorphically quantize the fields $f_{i}$ and $g_{i}$. $V_{D}$ and $V_{F}$ vanishes and the holomorphic fields does not depend on the angles of $S^{2}$ because we imposed spherical symmetry. We are left with
\begin{eqnarray}
n\int d\Omega dt \Big(\dot{f}_{1}^{\dag}\dot{f}_{1}+\dot{f}_{2}^{\dag}\dot{f}_{2}+\dot{g}_{1}^{\dag}\dot{g}_{1}+\dot{g}_{2}^{\dag}\dot{g}_{2}+\frac{1}{4}f_{1}^{\dag}f_{1}+\frac{1}{4}f_{2}^{\dag}f_{2}+\frac{1}{4}g_{1}^{\dag}g_{1}+\frac{1}{4}g_{2}^{\dag}g_{2}\Big),
\end{eqnarray}
the BPS equations implies
\begin{eqnarray}
\dot{f}_{i}=i\frac{1}{2}f_{i}\qquad  \dot{g}_{i}=i\frac{1}{2}g_{i}
\end{eqnarray}
therefore
\begin{eqnarray}\label{momentabps}
\Pi_{f_{i}}=-in\frac{1}{2}f^{\dag}_{i}\qquad \Pi_{g_{i}}=-in\frac{1}{2}g^{\dag}_{i}.
\end{eqnarray}
Now, let look at the CS action. After orbifold projection the gauge field is given by $A_{\mu}\otimes \mathbf{1}_{n\times n}$ and so, taking the trace is equivalent to do the rescaling $\kappa\rightarrow dim(R_{l})\kappa$ with $dim(R_{l})$ the dimension of the simple module in the node $l$. In our case there are only two nodes, both with $dim(R_{l})=n$. The equations of motion of $A_{\mu}$ on the sphere are automatically satisfied and we only have to look at the constraint from $A_{0}$. In this case, spherical symmetry imposes that all the fluxes must be equal hence the equations of $A_{0}$ and of $\hat{A}_{0}$ are equivalent and given by
\begin{eqnarray}
\frac{n\kappa}{\pi}\Phi&=&\int_{S^{2}} d\Omega\Big(|g_{1}|^{2}-|f_{2}|^{2}+|g_{2}|^{2}-|f_{1}|^{2}\Big)\nonumber\\
&=&\frac{2i}{n}\int_{S^{2}} d\Omega\Big(g_{1}\Pi_{g_{1}}+g_{2}\Pi_{g_{2}}-f_{1}\Pi_{f_{1}}-f_{2}\Pi_{f_{2}}\Big).
\end{eqnarray}
Note that this is a scalar equation now. We have to be careful when solving these equations and impose the quantization condition (\ref{quantcond}) for the fluxes. In this particular case we have, generically, a gauge group $G=U(nM)\times U(nN)$ or $G=U(M)\times U(N)$ in the locus where fractional branes are allowed. Let begin by taking the former case with $M\neq N$, $N<M$ therefore $n(M-N)$ components of the flux vanish and we have then $\Phi^{(1)}_{s}=\Phi^{(2)}_{s}=\Phi_{s}\otimes\mathbf{1}_{n\times n}$ with $\Phi_{s}\in 2\pi \mathbb{Z}$ for $s=1,\ldots, N$ labeling the different M2-branes. In order to have a well defined path integral we need, for each M2-brane (dropping the subindex $s$)
\begin{eqnarray}\label{quantflux}
\frac{n\kappa}{4\pi}(4\pi\Phi)\in  2\pi \mathbb{Z}\Rightarrow n\kappa\in \mathbb{Z}
\end{eqnarray}
so, the level $\kappa$ can be fractional. On the other hand, from the algebra point of view, this is implemented as follows. The coordinate ring that describe our moduli space will be given by polynomials of the form
\begin{eqnarray}\label{monomials}
f_{1}^{i_{1}}f_{2}^{i_{2}}g_{1}^{j_{1}}g_{2}^{j_{2}},
\end{eqnarray}
identifying $i\int_{S^{2}} d\Omega f\Pi_{f}$ with the number operator of the quantized theory, the monopole quantization condition imposes that the powers are related by
\begin{eqnarray}
i_{1}+i_{2}-j_{1}-j_{2}\in \mathbb{Z}n^{2}\kappa,
\end{eqnarray}
so define $\kappa=\frac{k}{n}$ with $k\in \mathbb{Z}$. Then, taking into account the action of the orbifold (\ref{actionorb}), the moduli space for a pointlike brane in the bulk is given by
\begin{eqnarray}
\mathcal{M}_{3d}=Sym^{N}(\mathbb{C}^{4}/(\mathbb{Z}_{nk}\times\mathbb{Z}_{n}\times\mathbb{Z}_{n}))
\end{eqnarray}
where $\mathbb{Z}_{nk}$ acts in the coordinates $\mathbb{C}^{4}$ as
\begin{eqnarray}
(f_{1},f_{2},g_{1},g_{2})\mapsto (e^{\frac{2\pi i}{kn}}f_{1},e^{\frac{2\pi i}{kn}} f_{2},e^{-\frac{2\pi i}{kn}}g_{1},e^{-\frac{2\pi i}{kn}} g_{2})
\end{eqnarray}
There are a couple of things to point out about this result. First, note that the $\mathbb{Z}_{nk}$ action is the remaining gauge symmetry from the $U(1)$ action we called $\lambda$ due to the periodicity of the monopole operator. Also we see that the action on the $S^{1}$ fiber is not $\mathbb{Z}_{n^{2}k}$ as we may have guessed at a first glimpse. Indeed, as we get before in our general analysis, the radius of the $S^{1}$ is further divided by the order of $q$ which in this case is $n$, the exponent of $H^{2}(\mathbb{Z}_{n}\times \mathbb{Z}_{n}, U(1))\cong \mathbb{Z}_{n}$.\\
The case $G=U(nN)\times U(nN)$ allows $\Phi_{s}=2\pi(m_{s}+\frac{p}{q})$ to be fractional. We will postpone the analysis of $\mathcal{M}_{3d}$ in these cases for future work, since it will require a more careful look at the dual operators in string/M- theory that carries the fractional charges.

\subsection{Singularities and fractional branes}

In this section we will focus on the singular locus where we expect to see brane fractionation. For this purpose we will first compute $\mathcal{ZA}_{q}$, since most of the singularities lay there, in the base of our moduli space. To be more precise, $\mathcal{ZA}_{q}$ will give us a commutative geometry for such $\mathcal{M}_{3d}$ is a $\mathbb{C}$ fibration over it. The variety $\mathcal{ZA}_{q}$ is what we expect from the moduli space of $D3$-branes on this singularity with discrete torsion. First note that the algebra $\mathcal{A}_{q}$ is Morita equivalent to the crossed product algebra $\mathcal{A}=\mathcal{A}_{c}\boxtimes(\mathbb{Z}_{n}\times \mathbb{Z}_{n})_{q}$ (see appendix (\ref{app:morita})) where the $q$ subindex indicate that $\mathbb{Z}_{n}\times \mathbb{Z}_{n}$ is twisted by the maximal cocycle $q$ and $\mathcal{A}_{c}$ is the conifold algebra spanned by $a_{i},b_{i}$.\\

\begin{figure}[h]
\centering
\includegraphics[height=5cm]{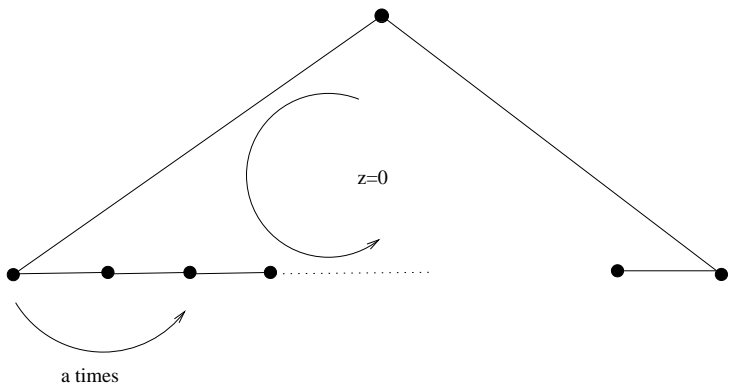}
\caption{Quiver diagram for the singularities at $z=0$. The nodes represent the $R_{l}$ modules }
\label{Fig.1}
\end{figure}

Two algebras that are Morita equivalent share the same center, so $\mathcal{ZA}_{q}$ is isomorphic to the ring of $\mathbb{Z}_{n}\times \mathbb{Z}_{n}$ invariants of $\mathcal{A}$. These are generated by
\begin{eqnarray}
u&=&(a_{1}b_{1})^{n}+(b_{1}a_{1})^{n},\nonumber\\
v&=&(a_{2}b_{2})^{n}+(b_{2}a_{2})^{n},\nonumber\\
w&=&(a_{1}b_{2})^{n}+(b_{2}a_{1})^{n},\nonumber\\
z&=&(a_{2}b_{1})^{n}+(b_{1}a_{2})^{n},\nonumber\\
t&=&a_{1}b_{1}a_{2}b_{2}+b_{1}a_{2}b_{2}a_{1},
\end{eqnarray}
therefore
\begin{eqnarray}
\mathcal{ZA}_{q}=\langle u,v,w,z,t\rangle/\{t^{n}-uv,t^{n}-zw\},
\end{eqnarray}
this describes the base of $\mathcal{M}_{3d}$, say a $\mathbb{Z}_{n}\times \mathbb{Z}_{n}$ orbifold of the conifold. Translating these variables to our original $A_{i}$, $B_{i}$ is direct, so we can read the base and the fibers in terms of the parameters $\{f_{i},g_{i}\}$. The singular locus of the base is given by the lines where any three of the $u,v,w,z$ variables are zero. The $\mathbb{Z}_{kn}$ action on the coordinates of the base and on the $S^{1}$ is free if we exclude the origin, so these are the only singularities we are getting. Let begin analyzing these singular lines, and postpone the analysis of the origin for later. All the singular lines are equivalent so, we can just focus in one particular case, say $z\neq 0$, then $f_{2}=g_{1}=0$. At these points, the modules we found above become reducible and so, a single point in the line corresponds to a family of $n$ $(1,1)$-dimensional simple modules that we denote by $R_{l}=R(f_{1}q^{l},g_{2},0,0)$, therefore we can write
\begin{eqnarray}
\lim_{u,v,w\rightarrow 0}R(f_{1},g_{2},f_{2},g_{1})=\oplus_{l}R_{l}.
\end{eqnarray}
These modules should be interpreted as fractional branes \cite{Berenstein:2001jr}. As we pointed out before, the allowed monomials are
\begin{eqnarray}
f_{1}^{i_{1}}g_{2}^{i_{2}}\qquad i_{1}-i_{2}\in \mathbb{Z}k
\end{eqnarray}
as long as $G=U(N)\times U(M)$ with $M\neq N$. Locally the singularity looks like $\mathbb{C}^{2}/\mathbb{Z}_{n}\times \mathbb{Z}_{k}$. The correct way to interpret this singularity is as a $k$-sheeted cover of $A_{n-1}$ singularities, that is $\left(\mathbb{C}^{2}/\mathbb{Z}_{n}\right)/\mathbb{Z}_{k}$. At fixed $f_{1}$ and $g_{2}$ each of these fractional branes can be associated with a twisted sector of the closed strings that is stuck at that point, which couples to the brane. Closed string twisted sectors (from the type IIA point of view) corresponds to operators of the chiral ring of the form $Tr((A_{i}B_{i})^{l})$. Is easy to see that these operators vanish everywhere in the bulk, for $n\nmid l$, since the trace gives a sum of characters. Only on the singularities they are non-zero and we have exactly $n-1$ classes of them that are trivial in the bulk. For example, in the particular locus we are looking at, the $\mathcal{O}_{\tau_{2}^{-l}}=Tr((A_{2}B_{1})^{l})$ operators are in the $\tau_{2}^{-l}$ twisted sector, they transform by a phase of $\varepsilon(g,\tau_{2}^{-l})=q^{-l}$ under the action of $\mathbb{Z}_{n}\times\mathbb{Z}_{n}$. These are the deformation modes that give us the necessary parameters to resolve the $A_{n-1}$ singularities (locally) in a bouquet of $n$ $2$-spheres. This basis of twisted sectors is related by a discrete Fourier transform to the basis of $\mathbb{CP}^{1}$'s of the blown up cycles. The discrete torsion generates a monodromy on this basis \cite{Gomis:2000ej}, as we go around the singularity. This can be seen directly from our computation. A loop around the singularity, $z\rightarrow ze^{2\pi i}$ is equivalent to $f_{1}\rightarrow f_{1}q^{a}$, where $a$ is the integer solution of $am+k'n=1$ with $k'\in\mathbb{Z}$  which means $R_{l}\rightarrow R_{l+a}$ when we go around a loop (Fig. \ref{Fig.1}).\\
We also see that the M-theory fiber is $n$ times larger that in the bulk, so the momentum along the circle is quantized in a different way. This effect takes into account the existence of monopoles of fractional charge, that can be written schematically as $f_{1}^{mk}g_{2}^{m'k}$. Note that this is exactly what we expected, from the analysis in section \ref{sec:regm2}, since the subgroup $H=\mathbb{Z}_{n}$ has trivial discrete torsion, hence $dim(H_{i})=1$\\

Now, let examine more carefully the points at the origin of the base $z=w=u=v=0$. At these points either $f_{i}\neq 0$ or $g_{i}\neq 0$, but only one of the four coordinates describing $\mathcal{M}_{3d}$ in non zero (if we are on the case that all of them are zero, we get back the theory we began with, as expected). Then a subgroup $\mathbb{Z}_{n}\times \mathbb{Z}_{n}$ of the total orbifold will left the point fixed. That means locally we have a $\mathbb{C}^{3}/(\mathbb{Z}_{n}\times \mathbb{Z}_{n})_{q}$ where the subindex $q$ indicates the discrete torsion. Likewise the case analyzed in \cite{Witten97} the moduli space is given by $nk$ copies of this singularity, say
\begin{eqnarray}
\left(\mathbb{C}^{3}/(\mathbb{Z}_{n}\times \mathbb{Z}_{n})_{q}\right)/\mathbb{Z}_{nk}
\end{eqnarray}

\section{Example 2: Orbifold by $\mathbb{Z}_{n}\times\mathbb{Z}_{m}$ with discrete torsion}

\begin{figure}[h]
\centering
\includegraphics[height=5cm]{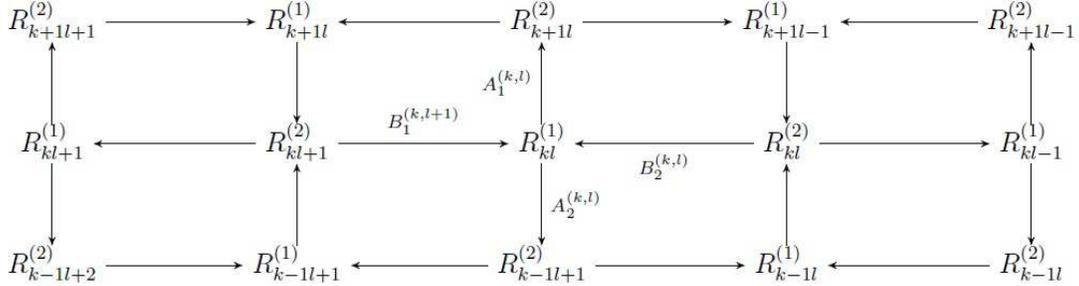}
\caption{Quiver diagram for $\mathbb{Z}_{n}\times\mathbb{Z}_{m}$ orbifold}
\label{Fig.3}
\end{figure}

Now consider an orbifold by $\Gamma=\mathbb{Z}_{n}\times\mathbb{Z}_{m} = \langle \tau_{1}\rangle\times \langle \tau_{2}\rangle$, acting on the bifundamental fields as
\begin{eqnarray}\label{actionorbmn}
\tau_{1}:(A_{1},A_{2},B_{1},B_{2})\mapsto (\alpha A_{1},\alpha^{-1} A_{2},B_{1},B_{2})\nonumber\\
\tau_{2}:(A_{1},A_{2},B_{1},B_{2})\mapsto ( A_{1},\beta A_{2},\beta^{-1}  B_{1},B_{2})
\end{eqnarray}
where $\alpha=e^{\frac{2\pi i}{n}}$ and $\beta=e^{\frac{2\pi i}{m}}$. Define $p=gcd(m,n)$ and so the discrete torsion of $\Gamma$ will be determined by an element of $H^{2}(\Gamma,U(1))\cong\mathbb{Z}_{p}$. Consider in general a cocycle $\eta=e^{\frac{2\pi ir}{p} }$ with $0\leq r< p$ and define $s$ as the smallest positive integer such that $\eta^{s}=1$. Then an irreducible representation is given by the $s\times s$ matrices
\begin{equation}\label{pqmatrices}
\gamma(\tau_{1})=P=%
\begin{bmatrix}
1 & 0 & 0 &\cdots & 0 \\
0 & \eta^{-1} & 0 & \cdots & 0 \\
0 & 0 & \eta^{-2} & \cdots & 0 \\
0 & 0 & 0 & \ddots & 0 \\
0 & 0 & 0 & \cdots & \eta^{s-1} %
\end{bmatrix}%
\text{, \ }\gamma(\tau_{2})=Q=%
\begin{bmatrix}
0 & 1 & 0 &\cdots & 0 \\
0 & 0 & 1 & \cdots & 0 \\
0 & 0 & 0 & \ddots & 0 \\
0 & 0 & 0 & \cdots & 1 \\
1 & 0 & 0 & \cdots & 0 %
\end{bmatrix}%
\text{,}
\end{equation}
so
\begin{equation}
\gamma(\tau_{1}^{a}\tau_{2}^{b})=P^{a}Q^{b}\qquad\alpha(\tau_{1}^{a}\tau_{2}^{b},\tau_{1}^{c}\tau_{2}^{d})=\eta^{bc}
\end{equation}
Any other non (linearly) equivalent irreducible projective representation can be written as $R_{kl}=span(\alpha^{k}\gamma(e_{1}),\beta^{l}\gamma(e_{2}))$ with $0\leq k< \frac{m}{s}$ and $0\leq l< \frac{n}{s}$. All these representations are of dimension $s$. The field theory associated with this orbifold has $2\frac{mn}{s^{2}}$ gauge groups and its quiver can be written over a torus. Here we draw a piece of it (Fig.\ref{Fig.3})
however, the superpotential is not given by the usual clockwise minus anti-clockwise squares as in an orbifold without discrete torsion. There exist relative phases between the squares we should determine. In order to do this we have to solve the equations (\ref{orbconseq}) for the fields $A_{i}, B_{i}$ when $\gamma$ is the regular representation
\begin{equation}
\bigoplus_{0\leq k<\frac{n}{s},0\leq l<\frac{m}{s}}s R_{kl}
\end{equation}
so the generators are given by
\begin{equation}
\gamma_{R}(\tau_{1})=\bigoplus_{i=0}^{\frac{m}{s}-1}\alpha^{i}P^{\oplus n}\qquad \gamma_{R}(\tau_{2})=\left(\bigoplus_{i=0}^{\frac{n}{s}-1}\beta^{i}Q^{\oplus s}\right)^{\oplus \frac{m}{s}}.
\end{equation}
The solutions to (\ref{orbconseq}) can be written in terms of the matrices
\begin{equation}
M_{1}=%
\begin{bmatrix}
0 & \mathbf{1}_{ns\times ns} & 0 &\cdots & 0 \\
0 & 0 &  \mathbf{1}_{ns\times ns} & \cdots & 0 \\
0 & 0 & 0 & \ddots & 0 \\
0 & 0 & 0 & \cdots &  \mathbf{1}_{ns\times ns} \\
(Q^{a})^{\oplus n} & 0 & 0 & \cdots & 0 %
\end{bmatrix}%
\text{, \ }M_{2}=\mathbf{1}_{\frac{m}{s}\times \frac{m}{s}}\otimes%
\begin{bmatrix}
0 & \mathbf{1}_{s^{2}\times s^{2}} & 0 &\cdots & 0 \\
0 & 0 &  \mathbf{1}_{s^{2}\times s^{2}} & \cdots & 0 \\
0 & 0 & 0 & \ddots & 0 \\
0 & 0 & 0 & \cdots &  \mathbf{1}_{s^{2}\times s^{2}} \\
(P^{a})^{\oplus s} & 0 & 0 & \cdots & 0 %
\end{bmatrix}\nonumber
\text{,}
\end{equation}
both elements of $M_{mn}(\mathbb{C})$ and invertible
\begin{equation}
M_{1}^{-1}=%
\begin{bmatrix}
0 &0 & 0 &\cdots & (Q^{-a})^{\oplus n} \\
 \mathbf{1}_{ns\times ns} & 0 &  0 & \cdots & 0 \\
0 &  \mathbf{1}_{ns\times ns} & 0 & \cdots & 0 \\
0 & 0 & \ddots & \cdots & 0 \\
0 & 0 & 0 &  \mathbf{1}_{ns\times ns} &  0 %
\end{bmatrix}%
\text{, \ }M_{2}^{-1}=\mathbf{1}_{\frac{m}{s}\times \frac{m}{s}}\otimes%
\begin{bmatrix}
0 &0 & 0 &\cdots & (P^{-a})^{\oplus s} \\
\mathbf{1}_{s^{2}\times s^{2}} & 0 &  0 & \cdots & 0 \\
0 &  \mathbf{1}_{s^{2}\times s^{2}} & 0 & \cdots & 0 \\
0 & 0 & \ddots & \cdots & 0 \\
0 & 0 & 0 &  \mathbf{1}_{s^{2}\times s^{2}}&  0 %
\end{bmatrix}\nonumber
\end{equation}
where $a$ is chosen to satisfy $-\frac{ar}{p}+\frac{1}{s}\in\mathbb{Z}$, equation that always have a solution, indeed, if $r\mid p$ we can chose $a=1$, if $r\nmid p$ then we have to solve $p k'+ra=1$ for $a,k'\in\mathbb{Z}$. $M_{1}$ and $M_{2}$ satisfy the relations
\begin{eqnarray}
\gamma_{R}(g_{1})M_{1}\gamma_{R}(e_{1})^{-1}&=&\alpha^{-1}M_{1}\nonumber\\
\gamma_{R}(g_{1})M_{2}\gamma_{R}(e_{1})^{-1}&=&M_{2}\nonumber\\
\gamma_{R}(g_{2})M_{1}\gamma_{R}(e_{2})^{-1}&=&M_{1}\nonumber\\
\gamma_{R}(g_{2})M_{2}\gamma_{R}(e_{2})^{-1}&=&\beta^{-1}M_{2}
\end{eqnarray}
So we can write the fields, after orbifold projection as
\begin{equation}
A_{1}=%
\begin{bmatrix}
0 & \cdots & 0 & 0 & A_{1}^{(\frac{m}{s}-1,0)}\otimes Q^{-a} &\cdots & 0 \\
0 & \cdots & 0 & 0 & 0 & \ddots & 0 \\
0 & \cdots & 0 & 0 & 0 &\cdots & A_{1}^{(\frac{m}{s}-1,\frac{n}{s}-1)}\otimes Q^{-a} \\
A_{1}^{(0,0)}\otimes \mathbf{1}_{s\times s} & \cdots & 0 & 0 & 0 & \cdots & 0 \\
0 & A_{1}^{(0,1)}\otimes \mathbf{1}_{s\times s} & 0 & 0 & 0 & \cdots & 0 \\
0 & 0 & \ddots & 0 & 0 & \cdots & 0 \\
0 & 0 & 0 &A_{1}^{(\frac{m}{s}-2,\frac{n}{s}-1)}\otimes \mathbf{1}_{s\times s}  & 0 & 0 & 0
\end{bmatrix}\nonumber
\end{equation}
the other variables can be projected likewise, $A_{1}\sim M_{1}^{-1}$, $A_{2}\sim M_{1}M_{2}^{-1}$, $B_{1}\sim M_{2}$, $B_{2}\sim 1$. Then the superpotential will have the form of a square superpotential algebra but one of the anti-clockwise squares will be multiplied by a phase $\eta^{a^{2}}$. We cannot eliminate this phase just by a redefinition of the fields since the following constraint must be satisfied
\begin{eqnarray}\label{prodbox}
\frac{\prod_{\Box_{C}}\lambda_{C}}{\prod_{\Box_{A}}\lambda_{A}}=\eta^{a^{2}}
\end{eqnarray}
where $\Box_{C}$ and $\Box_{A}$ stand for clockwise and anti-clockwise squares respectively. The quantity (\ref{prodbox}) is invariant under field redefinitions by multiplication of a phase. However, we can opt for a more symmetric choice and have all the anti-clockwise squares rescaled by
\begin{eqnarray}
\lambda^{s}\equiv e^{\frac{2\pi i a s}{mn}}
\end{eqnarray}
we define the projectors $P^{(i)}_{kl}$ associated with the gauge groups $R^{(i)}_{kl}$ as usual and the $A_{i}=\sum_{kl}A_{i}^{(kl)}$ and $B_{i}$ variables likewise. We also define the monomials
\begin{eqnarray}
\sigma_{k}^{(i)}=\sum_{j=1}^{\frac{n}{s}}\delta^{j}P^{(i)}_{kj}\qquad \delta=\beta^{s}
\end{eqnarray}
Then the relations of the generators of the path algebra are derived from the superpotential
\begin{eqnarray}
W=Tr\left[A_{1}B_{1}A_{2}B_{2}-\lambda^{s}A_{1}B_{2}A_{2}B_{1}\right]
\end{eqnarray}
plus
\begin{eqnarray}
\sigma^{(2)}_{k}B_{1}=\delta B_{1} \sigma^{(1)}_{k}\qquad \sigma^{(2)}_{k}B_{2}=B_{2} \sigma^{(1)}_{k}\qquad \sigma^{(1)}_{k}A_{2}=\delta A_{2} \sigma^{(2)}_{k-1}\qquad \sigma^{(1)}_{k}A_{1}= A_{1} \sigma^{(2)}_{k+1}\nonumber
\end{eqnarray}
Call this path algebra $\mathbb{C}Q_{\eta}$. Solving the modules of this path algebra is not an easy task, however, we can solve the problem for an algebra we know much better how to handle, the crossed product $\mathcal{A}=\mathcal{A}_{c}\boxtimes(\mathbb{Z}_{n}\times\mathbb{Z}_{m})_{\eta}$. These algebras are Morita equivalent (see appendix \ref{app:morita}). This fact allow us to extend our previous results in a rather simple way. The CS levels are rescaled by a factor of $s$ after we take the trace over the simple modules $R^{(a)}_{kl}$. Therefore, the quantization of the levels is taken to be $\kappa^{(a)}_{kl}=\frac{k}{s}$ with $k\in \mathbb{Z}$. Morita equivalence implies that the simple modules of $\mathbb{C}Q_{\eta}$ are again parametrized by four complex variables with a canonical kinetic term rescaled by a factor $\frac{mn}{s}\mathcal{S}_{kin}(f_{i},g_{i})$ \footnote{The $U_{\chi}$ matrices we define in section \ref{sec:regm2} will be of dimension $s$ and there are $\frac{mn}{s^{2}}$ irreducible representations. This is also clear from the result in the appendix \ref{app:morita}, where it can be seen directly from (\ref{moreq}) that the dimension of the simple $\mathbb{C}Q_{\eta}$-modules is $\frac{mn}{s}$ }. These two facts together gives a total moduli space in the bulk of the form
\begin{eqnarray}
\mathcal{M}_{3d}=\mathbb{C}^{4}/(\mathbb{Z}_{n}\times\mathbb{Z}_{m}\times\mathbb{Z}_{\frac{mnk}{s}})
\end{eqnarray}
which is the result we expected from our general analysis. The singularities can be derived as before from $\mathcal{ZA}$, which is given by (using the notation of appendix \ref{app:conifold})
\begin{eqnarray}
\mathcal{ZA}\cong\langle X_{1}^{\frac{mn}{p}},X_{2}^{\frac{mn}{p}},Y_{1}^{n},Y^{n}_{2},X_{1}X_{2}, (X_{1}Y_{2})^{m},(X_{2}Y_{1})^{m}\rangle/\{X_{1}X_{2}-Y_{1}Y_{2}\}
\end{eqnarray}
As we can see from the action of $\Gamma$, the fixed lines are the same as in the case of $\mathcal{A}_{q}$ therefore the analysis can be carried on on a similar way, likewise in \cite{Berenstein:2000hy}. The singularities are also locally $A_{s-1}$ but we will have a different factor on the monodromy of the $2$-spheres. Obviously, when $m=n$ we reproduce the result from section \ref{sec:qdefsec}.

\section{Matching with gravity duals}

The theories we have studied differ from their counterparts with trivial discrete torsion only by the superpotential. These deformations are casted by turning on a non-trivial NS $B$-field and RR fluxes. In general, as shown in \cite{Connes:1997cr,Seiberg:1999vs}, when we have a non-zero $B$-field it sources a non-commutativity parameter for the open strings. When we look at the worldvolume theory of a brane in these setups, the usual pointwise product between fields should be replaced by a star product. Since the non-commutativity parameter is controlled by the $B$-field, in the case it only has components transverse to the brane, the star product is given by constant phases
$$
f\star g=e^{i\pi\gamma(Q_{1}(f)Q_{2}(g)-Q_{2}(f)Q_{1}(g))}fg
$$
where the $Q_{i}$'s correspond to global charges and $\gamma$ is the non-commutativity parameter. The geometries dual to these deformed theories have been studied as well. For example, in the $AdS_{5}/CFT_{4}$ case, $\mathcal{N}=4$ SYM with a $q$-deformed superpotential corresponds to a near horizon geometry with a deformed 5-sphere $AdS_{5}\times\widetilde{S}^{5}$ \cite{Lunin:2005jy}. This metric can be generated by  a so called $TsT$ transformation, a solution generating technique presented in \cite{Lunin:2005jy}. The geometry with $\widetilde{S}^{5}$ is related by mirror symmetry to the orbifolded 5-sphere \cite{Berenstein:2000ux}.\\

In general, a $TsT$ transformation can always be realized, as long as our geometry possesses a $T^{2}$ fibration. The metric dual to ABJM has enough symmetry to do this. Their $q$-deformed counterparts have been actually computed \cite{Imeroni:2008cr}. The near horizon geometries we got from our analysis must be of the form $AdS_{4}\times \mathbb{C}P^{3}/\mathbb{Z}_{n}\times\mathbb{Z}_{n}$ in the $q$-deformed case (or $AdS_{4}\times S^{7}/(\mathbb{Z}_{n}\times\mathbb{Z}_{n}\times\mathbb{Z}_{nk})$ in M-theory). These seem quite different from the deformed backgrounds obtained in \cite{Imeroni:2008cr}, but there is no contradiction. Like in the type IIB case \cite{Dasgupta:2000hn,Berenstein:2000hy,Berenstein:2000ux} they are related by mirror symmetry. Let us see how this works. The moduli spaces we found can be seen as a $\mathbb{C}^{*}$-fibration over a basis described by $\mathcal{ZA}$, so we can use the same reasoning as in \cite{Dasgupta:2000hn,Berenstein:2000hy,Berenstein:2000ux}. From the M-theory point of view our bulk branes are expected to be M5-branes with topology $\mathbb{R}^{1,2}\times T^{2}\times S^{1}_{M}$, where $S^{1}_{M}$ is the extra M-theory circle. In type IIA string theory these are $D4$-branes with topology $\mathbb{R}^{1,2}\times T^{2}$. At the singularities this torus pinches $n$ times, giving us the fractional branes, which may then have topology $\mathbb{R}^{1,2}\times S^{2}\times S^{1}_{M}$. In type IIA these $D4$-branes have $D2$-brane charge due to Myers effect \cite{Myers:1999ps}. We can then use T-duality along the $T^{2}$ directions to interchange $D2/D4$ charges, therefore ending with a single point-like $D2$-brane and a fractional $B$-field. This is the geometry found in \cite{Imeroni:2008cr}. By an analysis of the DBI action of a single $D4$-brane they also found new branches when the deformation parameter $q$ is a root of unity, results that we confirm in our computation.  The interesting new effect we found is that the $S^{1}$ circle gets shrunk by a factor that depends on the choice of discrete torsion. The backgrounds from \cite{Imeroni:2008cr} do not present this effect, although the dilaton gets modified. It would be interesting to see if this effect could be reproduced from the SUGRA point of view. However, we may need to take into account corrections in $\alpha'$ of the background. One reason to think this is because the deformed $S^{7}/\mathbb{Z}_{k}$ metric has small curvature as long as $N^{2}k\gamma\ll 1$ (with $q=e^{2\pi i\gamma}$) then we need $n\gg 1$ if we want the 11d SUGRA approximation to be reliable. If we want to use the SUGRA approximation for $S^{7}/(\mathbb{Z}_{n}\times\mathbb{Z}_{n}\times\mathbb{Z}_{nk})$, then we need $n\ll 1$. \\

Is also interesting to note the resemblance with the moduli space obtained in \cite{Imamura:2008ji,Imamura:2008nn}, although the field theories are different, having vanishing CS levels for some of the hypermultiplets in the latter case. They obtained, by a brane construction, the dual theory on a singularity of the form $\mathbb{C}^{4}/\mathbb{Z}_{n}\times\mathbb{Z}_{m}\times\mathbb{Z}_{mn\kappa}$, so it would be worth studying how discrete torsion can be implemented there in order to see if the effect in the M-theory fiber is also reproduced in those theories and ultimately shed some light on the brane constructions that may lead to dual theories presented in this paper.

\section{Conclusions}

We analyzed the inclusion of discrete torsion on orbifolds of the ABJM-type family of theories. We focused on orbifolds by a group $\Gamma\leq SU(2)\times SU(2)$, in order to break supersymmetry at most to $\mathcal{N}=2$. We used the approach on \cite{Berenstein:2009ay} but using projective representations of $\Gamma$ which we showed is equivalent to twisting the crossed product algebra $\mathcal{A}_{c}\boxtimes \Gamma$. We found that, if we carefully take into account the fact that the smallest irreducible representation of $\Gamma$ is of order greater than one, we can have fractional CS levels on the covering theory. If we are twisting the algebra by a cocycle of order $m$, the CS levels in the covering can be chosen as $\kappa=\frac{k}{m}$ with $k$ relatively prime to $m$ giving us an orbifold theory with integer CS levels $k$. The net effect is that the smallest unit of D0-brane charge for the BPS monopoles is smaller than their counterparts without discrete torsion. This translates in a reduction of the radius of the M-theory circle by a factor of $\frac{m}{|\Gamma|k}$. This was shown very rigorously for the cases of $\Gamma$ abelian, but is expected to hold in general. At least the possibility of $\kappa$ being fractional was shown not to depend on this fact.\\

There are many possible future directions for this work. One of them is to consider deformations of more general quivers. However, for superpotentials of order different than four in the superfields, we will no longer have a canonical K\"ahler potential, making the holomorphic quantization procedure more difficult to implement. Another interesting question is what are the gravity duals of the theories we analyzed. We know the near horizon geometries found in \cite{Imeroni:2008cr} are good candidates, but it would be interesting to understand how T-duality acts on the full solution to give the correct value of the radius of the extra circle. This may shed some light also on what are the possible brane constructions that gives these configurations and ultimately a better understanding of the role of discrete torsion in M-theory.\\

Last but not least, we should mention that a careful study of the quantization conditions in the cases where fractional flux is allowed is worth studying. In particular, one of the important issues in these cases is to find what are the dual objects that carry these quantum numbers. We expect them to be baryon like operators as in the case analyzed in \cite{Berenstein:2009sa}.

\section*{Acknowledgements}
I would like to thank D. Berenstein for suggesting to me this project and for many helpful discussions and valuable advice. I thank C. Asplund, C. Beil, R. Eager and D. Morrison for various fruitful discussions and exchanges. In particular I would like to thank S. Franco for discussions, reading the manuscript and giving useful comments. I would also like to thank the Institute for Advanced Studies at Princeton for its hospitality during the last stages of this work. This research was supported in part by the DOE under grant DE-FG02-91ER40618.

\appendix

\section{Proof of theorem}\label{app:proof}

The results we will use to prove the theorem are in \cite{Karpilovsky,Humphreys78,Tappe77}. Let begin recalling that a theorem by Schur states that for every finite group $\Gamma$ there exists a \emph{representation group} $H$ and $A\leq Z(H)$, such that $\Gamma\cong H/A$ and $A\leq H'=[H,H]$ and $A\cong H^{2}(\Gamma,\mathbb{C}^{*})$. Then given an isomorphism $\theta:\Gamma\rightarrow H/A$ we can fix an element $r(g)\in H$ for each $g\in \Gamma$ such that $\theta(g)=Ar(g)$, so there exist a map $\beta:\Gamma\times \Gamma\rightarrow A$ that satisfies $r(g)r(g')=\beta(g,g')r(gg')$.\\

Now, consider a 2-cocycle $\alpha$ of $\Gamma$. We say that $\alpha$ is \emph{special} if given a character $\lambda$ of $A$ then $\alpha(g,g')=\lambda(\beta(g,g'))$ for all $g,g'\in \Gamma$. Then, we have two key results. First , there is an isomorphism $Hom(A,\mathbb{C}^{*})\cong H^{2}(\Gamma,\mathbb{C}^{*})$ induced by the map $\lambda\rightarrow [\lambda(\beta(g,g'))]$, so in every cocycle class of $\Gamma$ there exist a unique special 2-cocycle. Second, given a special special 2-cocycle $\alpha$ and $P_{i}$, $i=1,\ldots r$ representatives of the inequivalent projective representations of $\Gamma$ with cocycle $\alpha$, then there exists $r$ irreducible linear representations of $H$, $D_{i}$ such that $D_{i}(r(g))=P_{i}(g)$ for all $g\in \Gamma$. Now, proving our claim is very simple, consider
\begin{eqnarray}
D_{i}(r(g))D_{i}(r(g'))&=&\alpha(g,g')D_{i}(r(gg'))=\lambda(\beta(g,g'))D_{i}(r(gg'))\nonumber\\
&=&D_{i}(r(g)r(g'))=D_{i}(\beta(g,g'))D_{i}(r(gg'))
\end{eqnarray}
then $D_{i}(\beta(g,g'))=\lambda(\beta(g,g'))\mathbf{1}$ and taking the determinant of this expression we get ($det(D_{i}(a))=1$ for any $a\in A$ since $A\leq H'$)
\begin{eqnarray}
1=\lambda(\beta(g,g'))^{dim(P_{i})}\qquad \forall g,g'\in \Gamma
\end{eqnarray}
therefore if $m[\alpha]=0$ then $m\mid dim(P_{i})$, so the order of a given cocycle class divides the dimension of all its projective representations.

\section{Conifold algebra}\label{app:conifold}

The quiver that gives the conifold algebra $\mathcal{A}_{c}$ is given by two nodes ($P_{1}$ and $P_{2}$) and four arrows $A_{i}$, $B_{i}$ with $i=1,2$. The path algebra relations are given by
\begin{eqnarray}
P_{1}A_{i}=A_{i}P_{2}\qquad P_{2}A_{i}=A_{i}P_{1}=0, \\\nonumber
P_{1}B_{i}=B_{i}P_{2}=0\qquad P_{2}B_{i}=B_{i}P_{1},
\end{eqnarray}
plus the ones derived from the superpotential
\begin{eqnarray}\label{conifoldw}
W=\frac{1}{2}Tr(\epsilon^{ij}\epsilon^{kl}A_{i}B_{k}A_{j}B_{l})=Tr\left(A_{1}B_{1}A_{2}B_{2}-A_{1}B_{2}A_{2}B_{1}\right),
\end{eqnarray}
which are
\begin{eqnarray}
A_{1}B_{2}A_{2}=A_{2}B_{2}A_{1},\\\nonumber
A_{1}B_{1}A_{2}=A_{2}B_{1}A_{1},\\\nonumber
B_{1}A_{2}B_{2}=B_{2}A_{2}B_{1},\\\nonumber
B_{1}A_{1}B_{2}=B_{2}A_{1}B_{1},
\end{eqnarray}
the center $\mathcal{Z}\mathcal{A}_{c}$ is generated by $X_{1}=A_{1}B_{1}+B_{1}A_{1}$, $X_{2}=A_{2}B_{2}+B_{2}A_{2}$, $Y_{1}=A_{1}B_{2}+B_{2}A_{1}$ and $Y_{2}=A_{2}B_{1}+B_{1}A_{2}$ (see for example \cite{Berenstein:2001uv}) which satisfy the relation $X_{1}X_{2}=Y_{1}Y_{2}$.

\section{Proof of Morita equivalence}\label{app:morita}

\textbf{Claim 1:} The algebra $\mathcal{A}=\mathcal{A}_{c}\boxtimes(\mathbb{Z}_{n}\times \mathbb{Z}_{n})_{q}$ where the $q$ subindex indicate that $\mathbb{Z}_{n}\times \mathbb{Z}_{n}$ is twisted by the maximal cocycle $q$ and $\mathcal{A}_{c}$ is the conifold algebra spanned by $a_{i},b_{i}$ is Morita equivalent to $\mathcal{A}_{q}$.\\\\
\emph{Proof:}
Define the variables
\begin{eqnarray}
\widetilde{a_{2}}=a_{2}\rtimes 1\qquad \widetilde{a_{1}}=a_{1}\rtimes e_{\tau_{1}}^{-1}e^{-1}_{\tau_{2}}\qquad \widetilde{b_{2}}=b_{2}\rtimes e_{\tau_{1}}\qquad \widetilde{b_{1}}=b_{1}\rtimes e_{\tau_{2}}.
\end{eqnarray}
where $\mathbb{Z}_{n}\times \mathbb{Z}_{n}\cong\langle \tau_{1}\rangle\times\langle \tau_{2}\rangle$ and $e_{\tau_{i}}$ are the generators of $\mathbb{C}(\mathbb{Z}_{n}\times \mathbb{Z}_{n})_{q}$. Then $[\widetilde{a_{i}},e_{\tau_{j}}]=[\widetilde{b_{i}},e_{\tau_{j}}]=0$, therefore the crossed product algebra can be split in a tensor product
\begin{eqnarray}
\mathcal{A}_{c}\boxtimes(\mathbb{Z}_{n}\times \mathbb{Z}_{n})_{q}\cong \mathcal{A}_{q}[\widetilde{a_{i}}, \widetilde{b_{i}}]\otimes \mathcal{A}[e_{\tau_{1}},e_{\tau_{2}}]
\end{eqnarray}
since $\mathcal{A}[e_{\tau_{1}},e_{\tau_{2}}]$ have a unique irreducible representation, for $q$ maximal, then is isomorphic to the space of matrices of dimension $n\times n$ and so, we have that the category of modules of $\mathcal{A}$ and of $\mathcal{A}_{q}$ are equivalent, which is the definition of Morita equivalence.\\\\
\textbf{Claim 2:} The algebra $\mathbb{C}\mathcal{A}_{\eta}$ is Morita equivalent to $\mathcal{A}=\mathcal{A}_{c}\boxtimes(\mathbb{Z}_{n}\times\mathbb{Z}_{m})_{\eta}$.\\\\
\emph{Proof:} Let begin by constructing the projectors for the twisted algebra. These are given by sums over the generators corresponding to the $\alpha$-regular elements. Is easy to see that these elements are given by $\tau_{1}^{as}\tau_{2}^{a's}$ with $1\leq a\leq\frac{n}{s}$ and $1\leq a'\leq\frac{m}{s}$, hence the projectors will be given by
\begin{eqnarray}
P_{kl}=\frac{s^{2}}{mn}\sum_{a,a'}\alpha^{kas}\beta^{la's}e_{\tau_{1}^{as}\tau_{2}^{a's}}\qquad P_{kl}P_{k'l'}=\delta_{kk'}\delta_{ll'}
\end{eqnarray}
at this point we can see that the projectors $P^{(a)}_{kl}\equiv P_{a}\rtimes P_{kl} $ satisfy the expected relations with the generators of $\mathcal{A}_{c}$, derived from the action (\ref{actionorbmn}) of $\mathbb{Z}_{n}\times\mathbb{Z}_{m}$ on them
\begin{eqnarray}
P^{(1)}_{kl}a_{1}=a_{1}P^{(2)}_{k+1l}\qquad P^{(1)}_{kl}a_{2}=a_{2}P^{(2)}_{k-1l+1}\qquad P^{(2)}_{kl}b_{1}=b_{1}P^{(1)}_{kl-1}\qquad P_{kl}^{(2)}b_{2}=b_{2}P^{(1)}_{kl}\nonumber
\end{eqnarray}
our goal is to factor an algebra that has a unique irreducible representation. The obvious choice is the algebra $\mathcal{A}[\tilde{e}_{1},\tilde{e}_{2}]$ generated by $\tilde{e}_{1}=e_{\tau_{1}^{\frac{n}{s}}}$ and $\tilde{e}_{2}=e_{\tau_{2}^{\frac{m}{s}}}$. Then $\mathcal{A}[\tilde{e}_{1},\tilde{e}_{2}]\cong\mathbb{C}(\mathbb{Z}_{s}\times\mathbb{Z}_{s})_{\eta}$. Define the variables
\begin{eqnarray}
\widetilde{a_{1}}=a_{1}\rtimes  e_{\tau_{2}^{b}}\qquad \widetilde{a_{2}}=a_{2}\rtimes e_{\tau_{1}^{c}}e_{\tau_{2}^{-b}}\qquad \widetilde{b_{2}}=b_{2}\rtimes 1\qquad \widetilde{b_{1}}=b_{1}\rtimes e_{\tau_{1}^{-c}}.
\end{eqnarray}
with $b$ and $c$ integers chosen such that
\begin{eqnarray}
\eta^{c\frac{m}{s}}=\eta^{b\frac{n}{s}}=e^{\frac{2\pi i}{s}}
\end{eqnarray}
therefore they commute with $\tilde{e}_{1}$ and $\tilde{e}_{2}$. A direct computation give us
\begin{eqnarray}
\widetilde{a_{1}}\widetilde{b_{2}}\widetilde{a_{2}}\widetilde{b_{1}}=\eta^{cb}\alpha^{-c}\beta^{b}\widetilde{a_{1}}\widetilde{b_{1}}\widetilde{a_{2}}\widetilde{b_{2}}
\end{eqnarray}
and, with a fair amount of patience it can be shown that $b$ and $c$ can be chosen such that $\eta^{cb}\alpha^{-c}\beta^{b}=\lambda^{-s}$ therefore showing that
\begin{eqnarray}\label{moreq}
\mathcal{A}\cong\mathbb{C}\mathcal{A}_{\eta}\otimes\mathcal{A}[\tilde{e}_{1},\tilde{e}_{2}]
\end{eqnarray}
which is the statement that both algebras are Morita equivalent.


\end{document}